\documentclass{aa}

\usepackage[varg]{txfonts}
\usepackage{hyperref}
\usepackage[usenames]{color}
\hypersetup{colorlinks=true, linkcolor=black, citecolor=black, filecolor=blue, urlcolor=blue}
\usepackage{graphicx}
\usepackage{array}

\usepackage[position=top]{subfig}

\begin{document} 

\title{Cosmic ray feedback from supernovae in dwarf galaxies}

\titlerunning{CR feedback from SNe in dwarf galaxies}
\authorrunning{G. Dashyan \& Y. Dubois}

\author{Gohar Dashyan \and Yohan Dubois}

\offprints{Y. Dubois, \email{dubois@iap.fr}}

\institute{Sorbonne Universit\'e, UPMC-CNRS, UMR7095, Institut d'Astrophysique de Paris, F-75014 Paris, France}

\date{Received 18 July 2019 / Accepted 18 April 2020}

\abstract{The regulation of the baryonic content in dwarf galaxies is a long-standing problem. Supernovae (SNe) are supposed to play a key role in forming large-scale galactic winds by removing important amounts of gas from galaxies. SNe are efficient accelerators of non-thermal particles, so-called cosmic rays (CRs), which can substantially modify the dynamics of the gas and conditions to form large-scale galactic winds. We investigate how CR injection by SNe impacts the star formation and the formation of large-scale winds in dwarf galaxies, and whether it can produce galaxy star-formation rates (SFR) and wind properties closer to observations. We ran CR magneto-hydrodynamical simulations of dwarf galaxies at high resolution ($9 \, \rm pc$) with the adaptive mesh refinement code {\sc ramses}. Those disc galaxies are embedded in isolated halos of mass of $10^{10}$ and $10^{11} \, \rm M_{\odot}$, and CRs are injected by SNe. We included CR isotropic and anisotropic diffusion with various diffusion coefficients, CR radiative losses, and CR streaming. The injection of CR energy into the interstellar medium smooths out the highest gas densities, which reduces the SFR by a factor of 2--3. Mass outflow rates are significantly greater with CR diffusion, by 2 orders of magnitudes for the higher diffusion coefficients. Without diffusion and streaming, CRs are inefficient at generating winds. CR streaming alone allows for the formation of winds but which are too weak to match observations. The formation of galactic winds strongly depends on the diffusion coefficient: for low coefficients, CR energy stays confined in high density regions where CR energy losses are highest, and higher coefficients, which allow for a more efficient leaking of CRs out of dense gas, produce stronger winds. CR diffusion leads to colder and denser winds than without CRs, and brings outflow rates and mass loading factors much closer to observations.}


\maketitle

\section{Introduction}

It has by now been well established that feedback -- the processes by which star formation and supermassive black hole growth regulate themselves by deposition of energy and momentum in the ambient medium -- is a key mechanism to regulate the amount of baryons in galaxies~\citep[see][for recent reviews]{Somerville15,NaabOstriker17}. Additionally, outflows are ubiquitously observed in galaxies (see review by \citealt{Veilleux2005}) and moderate density regions of the Universe -- where stars do not form -- are metal-enriched, which suggests that metal-enriched winds can deposit metals into the intergalactic medium \citep[e.g.][]{Booth12}. However, the processes regulating star formation and driving winds in galaxies, as well as their relative importance, are still a topic of active investigation.

Due to their shallow potential wells and lack of galactic mergers during their evolution, dwarf galaxies (with halo virial mass of $M_{\rm vir} \leq 10^{11}\, \rm M_{\odot}$) can act as both important probes for cosmological structure formation, and as astrophysical test-beds for feedback processes. As the least massive and least luminous galaxies, they are both difficult to resolve in cosmological simulations and under-observed. Several puzzles surrounding dwarf galaxies, including the too-big-to-fail problem~\citep{TBTF11}, the missing satellites~\citep{klypin99,Moore99Sat}, and the cusp-core problems~\citep{Moore99CuspCore}, have posed important challenges to our understanding of galaxy formation and dark matter (DM) for DM-only models that can potentially be alleviated by a realistic treatment of baryonic feedback processes~\citep[e.g.][]{PG12,Zolotov12,Teyssier13,BZ14,Chan15,Onorbe15}.

Supernova (SN) feedback remains the commonly believed dominant feedback mechanism in low mass halos (e.g. \citealt{Dekel}). The problem is that it is partly implemented in terms of subgrid physics in numerical simulations, which prevents the capture of its exact nature and extent. Ideally, we would like to directly simulate the physics without tuning the parameters to observables, and rather use observations as powerful discriminants for theories. 
Recent work argues for the need for interstellar medium (ISM) physics beyond SN feedback alone to suppress star formation and generate outflows \citep{Rosdahl17,Smith18}. 
In addition, resolution is generally too low to capture the detailed evolution of SN explosions~\citep{Hu18}, which has produced a diversity of subgrid workarounds such as delayed cooling~\citep{Stinson06,Teyssier13}, kinetic winds~\citep{Springel03,DT08}, stochastic feedback~\citep{DallaVecchia12}, or momentum-driven approaches~\citep{Kimm15}, with diverse successes at producing efficient outflows~\cite[see][for a comparison of such models]{Rosdahl17}.
Feedback mechanisms other than SNe are now suspected to play a key role in shaping galactic winds and regulating the baryon content in galaxies~\citep{Hopkins14}. The attention has therefore turned to other processes such as stellar radiation \citep{Hopkins11,Agertz13,Aumer13,Rosdahl15,Emerick18}, the momentum transferred from resonantly scattered Lyman-$\alpha$ photons \citep{Kimm}, or cosmic rays (CRs) released in SNe~\citep{Wadepuhl11,Booth13,SalemBryan14disc}.

 CRs are particles accelerated to relativistic velocities at shock fronts of supernova remnants (SNR) (see \citealt{Caprioli15} for a recent review), active galactic nuclei (e.g. \citealt{Berezinsky2006}), winds from massive stars \citep{Bykov2014}, or structure formation shocks (e.g. \citealt{Miniati2001}). Observations of local SN remnants suggest that of the order of 10\% of the explosion energy is converted to CRs \citep{Morlino2012,Helder13}. In the solar neighbourhood, CRs have an observed energy density of $\sim 10^{-12}$ ergs cm$^{-3}$ \citep{Wefel87}, which is comparable to the magnetic, turbulent and vertical gravitational energies. That equipartition shows that CRs are important in the local dynamics of the disc. Recent observations suggest that CR energy density exceeds that seen in the Milky Way by orders of magnitude in the starbust galaxy M82 \citep{Veritas2009,Paglione2012}. The close correlation observed between the radio continuum and the far-infrared emission of spiral galaxies (e.g. \citealt{DeJong85}) hints at a correlation between star formation and CR energy density and therefore at a powerful self-regulation mechanism.
 
The protons of a few GeV, where most of CR energy density resides, are essentially collisionless. Their coupling with the gas is mediated by the ambient magnetic field and can be described by a fluid picture that macroscopically describes the energy and momentum exchange between CRs and thermal gas, mediated by the gyroresonant streaming instability \citep{Kulsrud2005,Zweibel2017}. CRs have several properties which make them a promising candidate for driving galactic winds: they provide an additional pressure gradient that helps lifting thermal gas; due to its relativistic nature, the CR component has a softer equation of state  ($\gamma_{\rm CR} = 4/3$) than the thermal component ($\gamma = 5/3$) and therefore suffers less losses upon adiabatic expansion; the CR energy has a longer cooling time than that of the thermal component; finally, CRs diffuse along magnetic field lines and thereby move relative to the thermal component and leak out from their production sites, that is from dense and efficiently cooling star-forming regions to more diffuse regions.

The role of CRs on galaxy evolution and their ability to drive winds have been a subject of interest in both analytical \citep{Ipavich1975,Breitschwerdt1991,Recchia16,Mao18} and numerical studies. In hydrodynamical simulations without magnetic field evolution, \cite{Uhlig12} found that CR streaming -- assuming a streaming velocity proportional to the sound speed -- drives powerful and sustained winds in galaxies with virial masses $M_{\rm vir}<10^{11}\, \rm{M}_{\odot}$. \cite{Booth13}, performing simulations of Small Magellanic Cloud and Milky Way sized disc galaxies, showed that the star-formation rate (SFR) was suppressed for both masses when accounting for isotropic CR diffusion (i.e. diffusion of CRs that is independent of the magnetic field lines). With similar assumptions on the propagation of CRs, \cite{SalemBryan14disc} found that CRs thicken gaseous disc, suppress the SFR and produce mass loading factors (i.e. the ratio of the outflow rate to the star-formation rate) above unity in a Milky Way mass galaxy. 

In order to include the anisotropic nature of CR diffusion into simulations -- that is the fact that CRs diffuse along magnetic field lines preferentially -- the evolution of magnetic fields would have to be modelled since CRs gyrates around magnetic field lines, CR diffusion and streaming are parallel to them, and the streaming velocity depends on the local Alfv\'en velocity, which depends on the amplitude of the magnetic field.
Using magneto-hydrodynamical (MHD) simulations, 
\cite{Hanasz13} demonstrated that even with anisotropic diffusion and CR feedback alone, efficient large-scale winds can be driven out of galaxies. 
\cite{Pakmor16} showed that winds develop significantly later with anisotropic diffusion compared to isotropic diffusion. In simulations of a stratified disc, \cite{Girichidis16} showed that including CRs thickens the galactic disc and leads to the formation of a warm and neutral galactic atmosphere providing a mass reservoir for galactic winds and outflows, as opposed to feedback without CRs that produces hot and ionised outflows. \cite{Ruszkowski17} included CR streaming in numerical galactic MHD simulations. They found that the inclusion of CR streaming and anisotropic diffusion can have a significant effect on wind launching and mass loading factors. They also found that the presence of moderately super-alfv\'enic CR streaming greatly enhances the efficiency of galactic wind driving. However \cite{Holguin18}, taking into account realistic streaming instability suppression in the presence of turbulence that results in higher streaming speeds, found that the mass loading factor drops significantly as the strength of turbulence increases. \cite{Butsky2018}, simulating a suite of isolated Milky Way-type disc galaxies, found that CR feedback suppresses star formation by supporting thermal gas against collapse, that models with CR transport drive strong outflows and reproduce the multiphase temperature and ionisation structure of the circumgalactic medium (CGM), but that the simulated CGM strongly depends on CR transport model. \cite{Girichidis18} studied the impact of CR-driven galactic outflows from a multiphase ISM, and found that simulations including CR have denser, colder and slower outflows. Performing  computations of resonant Lyman-$\alpha$ radiation transfer through snapshots of a suite of stratified disc simulations, \cite{Gronke18} found that the absence of CR feedback leads to spectra incompatible with observations due to the smoother neutral gas distribution of CR supported outflows. \cite{Jacob18} performed a parameter study varying the halo mass, the diffusion coefficient and the CR injection fraction and found that CRs are only able to drive winds in halos with masses $M_{\rm vir}< 10^{12}\, \rm{M}_{\odot}$ and that the mass loading factor drops with virial mass. \cite{Chan2018} presented isolated simulations of dwarf and Milky Way-type galaxies including CR feedback. They found that only constant isotropic diffusion coefficients of $3\times 10^{28-29}\, \rm cm^2\, s^{-1}$ reproduce observations of $\gamma$-ray emission. With  advection-only or streaming-only models (even when allowing for super-alfv\'enic streaming speeds) too large $\gamma$-ray luminosities are obtained. However, the effects of CRs on SFR and gas density distributions that they obtained are relatively modest.

In a cosmological context,~\cite{Jubelgas08} studied the impact of CR acceleration on structure formation shocks and showed that CRs can significantly reduce the star-formation efficiency of small galaxies. \cite{Wadepuhl11} performed hydrodynamical simulations of the formation of Milky Way-sized galaxies and found that simulations that include CRs match the faint-end of the galaxy luminosity function. \cite{SalemBryan14Cosmo} performed cosmological zoom-in simulations of a $10^{12} \rm{M}_{\odot}$ halo and found that CRs generate thin, extended discs with a significantly more realistic rotation curve. \cite{ChenBryan16} performed cosmological zoom in simulations of dwarf galaxies; their simulations retrieve the observed baryonic Tully-Fisher relation and produce smoother SFRs than the bursty SFRs obtained with a thermal component only, but lead to too many stars and to the presence of DM cusps. Recently, \cite{Hopkins2019} presented the study of a large suite of high-resolution cosmological zoom-in simulations. They found that CRs have relatively weak effects on dwarf galaxies at redshifts above $1-2$ for a diffusion coefficient, which matches observations of $\gamma$-ray emission.
Overall, those numerical studies that have included CR transport and studied wind properties agree that CRs increase the mass loading and the density of galactic winds.

In the present work, we study the impact of CR feedback released by SNe on galaxies and wind formation using high resolution MHD simulations of dwarf galaxies of halo mass of $10^{10} \,\rm{M}_{\odot}$ and $10^{11} \,\rm{M}_{\odot}$, and varying the physics of CRs (isotropic or anisotropic diffusion, with different values of the diffusion coefficient, and testing for the effect of the streaming instability).

This paper is structured as follows. In Section \ref{simulations and methods}, we describe the setup of our isolated galaxy disc simulations. In Section \ref{results}, we present the results of our analysis on the most massive dwarf galaxy, focusing on the suppression of star formation, the density distribution in the ISM, the generation of outflows, the distribution of CR energy and the evolution of the magnetic field. We also discuss how these results depend on the CR transport model. In Section \ref{G8/G9} we discuss the dependency of CR-driven winds on halo mass. In Section \ref{convergence}, we examine how these results converge with numerical resolution. In Section \ref{Binit} we test the dependency on the initial topology of the magnetic field. We discuss and summarise our results in Section \ref{Conclusion}.

\newcolumntype{P}[1]{>{\centering\arraybackslash}m{#1}}
\setlength{\tabcolsep}{5pt}
\renewcommand{\arraystretch}{2.0}
\begin{table*}
	\centering
	\caption{Simulation initial conditions and parameters for the disc galaxy modelled in this paper. The listed parameters are, from left to right: $M_{\rm halo}$: DM halo mass, $R_{\rm vir}$: halo virial radius (defined as the radius within which the DM density is 200 times the critical density at redshift zero),  $v_{\rm circ}$: circular velocity at the virial radius, $L_{\rm box}$: simulation box length, $M_{\rm disc}$: disc galaxy mass in baryons (stars+gas), $M_{\rm bulge}$: stellar bulge mass, $f_{\rm gas}$: disc gas fraction, $N_{\rm part}$: number of DM/stellar particles, $m_{*}$: mass of stellar particles formed during the simulations,  $\Delta_{\rm xmax}$: coarsest cell resolution,  $\Delta_{\rm xmin}$: finest cell resolution, $Z_{\rm disc}$: disc metallicity.}
	\label{tab1}
    \begin{tabular}{P{1.50cm}P{1.01cm}P{1.01cm}P{1.01cm}P{1.01cm}P{1.2cm}P{1.2cm}P{1.01cm}P{1.01cm}P{1.2cm}P{1.01cm}P{0.9cm}P{0.9cm}}
		\hline
		Galaxy acronym &$M_{\rm halo}$ [$\rm{M}_{\odot}$] & $R_{\rm vir}$ [kpc] & $v_{\rm circ}$ [$ \rm{km\,s}^{-1}$]  & $L_{\rm box}$  [kpc]   &  $M_{\rm disc}$ [$\rm{M}_{\odot}$]   & $M_{\rm bulge}$  [$\rm{M}_{\odot}$] & $f_{\rm gas}$  & $N_{\rm part}$  & $m_{*}$ \,\,[$\rm{M}_{\odot}$] & $\Delta x_{\rm max}$ [kpc] & $\Delta x_{\rm min}$ [pc] & $Z_{\rm disc}$ [$Z_{\odot}$]  \\
		\hline
		  G9 & $10^{11} $ & $89$ &  $65$  & $300$ & $3.5 \times 10^9$  &   $3.5 \times 10^8$ & $0.5$  &  $10^6$  &  $2.0 \times 10^3$ & $2.3$ & $9$ & 0.1 \\
		\hline
		 G8 &  $10^{10} $ & $41$ &  $30$  & $150$ & $3.4 \times 10^8$  &   $3.5 \times 10^7$ & $0.5$  &  $10^6$  &  $2.0 \times 10^3$ & $2.3$ & $9$ & 0.1 \\
		\hline
	\end{tabular}
\end{table*}

\section{Simulations and methods}
\label{simulations and methods}

Using the adaptive mesh refinement code {\sc ramses} \citep{Teyssier2002}, we ran idealised simulations of rotating isolated disc galaxies, consisting of gas and stars, embedded in a DM halo. The equations of ideal magneto-hydrodynamics (MHD)~\citep{Fromangetal06} are computed using the Harten-Lax-van Lear-Discontinuities Riemann solver from~\cite{HLLLD05} and the MinMod slope limiter to construct variables at cell interfaces from the cell-centred values. The induction equation evolving the magnetic field is solved using constrained transport on the adaptive grid~\citep{Teyssieretal06}, which guarantees that the divergence of the magnetic field is zero at machine precision. Here we provide a short description of the simulation setup as well as the physical modules used, described in detail in \cite{Rosdahl17}. We used the method introduced by \cite{DuboisCommercon16} for solving the anisotropic diffusion of CR energy, that is, we used an implicit finite-volume method in the {\sc ramses} code. We detail the CR magnetohydrodynamics in Section \ref{CRs}.

\subsection{Initial conditions}

The main parameters for the simulated galaxies and their host DM haloes are presented in Table \ref{tab1}. Similar simulation settings were used in \cite{Rosdahl15,Rosdahl17}. The initial conditions are generated with the {\sc makedisc} code~\citep{Springeletal05}. We focus most of our analysis on the higher mass galaxy that we name G9. The rotating isolated galaxy of G9 has a baryonic mass of $M_{\rm bar} = M_{\rm disc}+ M_{\rm bulge} = 3.8 \times 10^9 \, \rm{M}_{\odot}$, with an initial gas fraction of $f_{\rm gas} = 0.5$, and is hosted by a DM halo of mass $M_{\rm halo} = 10^{11} \, \rm{M}_{\odot}$. We also compare, at the same spatial resolution, a less detailed set of results for feedback models in a ten times less massive galaxy, that we name G8, with a baryonic mass $M_{\rm{bar}} = 3.8 \times 10^8\, \rm{M}_{\odot}$, $f_{\rm gas} = 0.5$ and $M_{\rm halo} = 10^{10} \,\rm{M}_{\odot}$. Each simulation is run for 250 Myr that is 2.5 orbital times at the scale radii of G8 and G9.

The DM halo follows the NFW density profile of \citet*{NFW}, with a concentration parameter $c = 10$ and spin parameter $\lambda = 0.04$~\citep{Bullocketal01}. The DM is modelled by $10^6$ collisionless particles, hence the galaxy halo has a DM mass resolution of $10^5\, \rm{M}_{\odot}$ for G9 and $10^4\, \rm{M}_{\odot}$ for G8. The initial disc consists of stars and gas, both set up with density profiles that are exponential in radius and Gaussian in height from the mid-plane (scale radius of $0.7$ kpc for the G8 galaxy and $1.5$ kpc for the G9 galaxy; scale height of one-tenth of the scale radius). The G8 and G9 galaxies contain a stellar bulge with masses and scale radii both one-tenth that of the disc. The initial stellar particle number is $1.1 \times 10^6$, a million of which are in the disc and the remainder in the bulge. The mass of the initial stellar particles is $1.7 \times 10^2 \,\rm{M}_{\odot}$ for G8 and $1.7 \times 10^3 \,\rm{M}_{\odot}$ for G9 (close to the masses of stellar particles formed during the simulation for G9, but ten times smaller for G8). While contributing to the dynamical evolution and gravitational potential of the rotating galaxy disc, the initial stellar particles do not explode as SNe.
The temperature of the gas discs is initialised to a uniform $T = 10^4$ K and the ISM metallicity $Z_{\rm disc}$ is set to 0.1 for the G8 and G9 galaxies. The CGM initially consists of a homogeneous hot and diffuse gas, with $n_{\rm H} = 10^{-6} \,\rm{cm}^{-3}$, $T = 10^6$ K and zero metallicity. The cutoff radii and heights for the gas discs are chosen to minimise the density contrast between the disc edges and the CGM. The square box widths for the G8 and G9 galaxies are 150 kpc and 300 kpc, respectively, and we use outflow (i.e. zero gradient) boundary conditions on all sides.

This particular choice of idealised setup for the galaxy and its CGM is a great simplification of the cosmological infall composed of anisotropic cold narrow streams of matter and a hot and diffuse corona, with respective mass fraction depending on mass and redshift~\citep[see e.g.][]{Dekel09}. It is feasible, alternatively, to study an idealised problem with an initially spherically-symmetric halo, where the collapse of the gaseous halo interferes with the propagation of winds~\cite[e.g.][]{Jacob18}, but where such idealised setups are also a very simplified representation. The advantage of such a simplified setup is to allow for the study of the launching of galactic winds without the complex interplay with an inhomogeneous cosmological infall, whose aspect is beyond the scope of this paper.

\subsection{Magnetic field}

The initial magnetic field vector $\vec{B}$ is obtained by defining an initial value of the magnetic potential vector, so that the reconstruction of $\vec{B}=\nabla \times \vec{A}$ ensures $\nabla\cdot\vec{B}=0$~\citep{DT10}.
A toroidal $\vec{B}$ field is obtained by setting the value of $\vec{A}$ to 
\begin{equation}
    \vec{A}=\frac{3}{2}B_{0}r_{0}\left(\frac{\rho}{\rho_0}\right)^{2/3}\vec{e}_{z}\, ,
\end{equation}
where $\vec{e}_{z}$ is the $z$-cartesian unit vector, and $\rho$ is the gas density profile as defined in the disc within the vertical and radial cutoffs, $\rho_0$ its normalisation ($\sim 15 \rm{cm}^{-3}$ for both G8 and G9) and $r_{0}$ its scale radius (1.5 kpc for G9 and 0.7 kpc for G8). $B_0$ is set to 1 $\mu$G.
With this parametrization, a toroidal magnetic field with only non-zero initial components $B_x$ and $B_y$ is produced with a density-dependency $\rho^{2/3}$ that mimics the effect of pure compression of magnetic field lines in the disc. 

In Section \ref{Binit}, we test the influence of the initial topology of the magnetic field by changing the initial topology from toroidal to poloidal.

\subsection{Adaptive refinement}

Each refinement level uses half the cell width of the next coarser level, starting at the box width at the first level. Our simulations start at level 7 in a box of size $L_{\rm box}=300  \, \rm kpc$ for G9 and $L_{\rm box}=150  \, \rm kpc$ for G8, corresponding to a coarse resolution of $128^3$ cells, and adaptively refine up to a maximum level 15 for G9 and level 14 for G8, corresponding to a minimum cell width of $\Delta x_{\rm min}=9 \,\rm pc$. A cell is refined (derefined) up to the maximum level (respectively down to level 7) if its total mass exceeds 8 times $10^3 \,\rm{M}_{\odot}$ (respectively is below $10^3 \,\rm{M}_{\odot}$), or if its width exceeds a quarter of the local Jeans length. In Section \ref{convergence}, we perform a convergence study for the G9 galaxy, where the finest level is set to 14.

\subsection{Gas cooling}

Gas can radiatively cool down to a minimum temperature of $10^3 \, \rm K$ (but further down due to adiabatic expansion) following \cite{Sutherland&Dopita93} for metal-dependant cooling rates above $10^4\,\rm K$, and \cite{Rosen&Bregman95} below. 
We started with an initial metallicity of $0.1 \,\rm Z_\odot$ that we kept constant (no release of metals by stellar evolution).
We included heating of diffuse gas from a redshift zero \cite{Haardt1996} ultraviolet (UV) background, and enforce an exponential damping of the UV radiation above the self-shielding density of $n_{\rm H}= 10^{-2}\,\rm{cm}^{-3}$.

\subsection{Star formation}

Our star formation model follows a Kennicutt-Schmidt law, $\dot{\rho}_*= \epsilon_*\rho/t_{\rm ff}$ \citep{Kennicutt98,Krumholz2007} where $\dot{\rho}_*$ is the SFR density, $\epsilon_* = 0.02$ the constant star-formation efficiency, $\rho$ the gas density and $t_{\rm ff}$ is the free fall time of the gas. Star particles of mass $m_*=2 \times 10^3 \rm{M}_{\odot}$ are formed stochastically using a Poissonian distribution (see \citealt{Rasera06} for details), in cells that exceed the star-formation number density threshold $n_0=80$ cm$^{-3}$ and with a temperature below $3\times 10^3 \,\rm K$. 

To prevent numerical fragmentation of gas below the Jeans scale \citep{Truelove97}, an artificial `Jeans pressure' is maintained in each gas cell in addition to the thermal pressure. In terms of an effective temperature, it can be written as $T_{\rm J} = T_0 n_{\rm H}/n_0$, where we have set $T_0 = 10^3$ K (and $n_0$ is the aforementioned star-formation threshold), to ensure that the Jeans length is resolved by a minimum number of 7 cell widths at any density (see equation 3 in \citealt{Rosdahl15}). The pressure floor is non-thermal, in the sense that the gas temperature that is evolved in the thermochemistry is the difference between the total temperature and the floor -- therefore, we can have $T \ll T_{\rm J}$.

\subsection{Supernova feedback}

SN feedback is performed with single and instantaneous injections per stellar population particle $t_{\rm SN} = 5 \,\rm{Myr}$ after it has formed. Each stellar particle has an energy ($E_{\rm SN}$) and mass ($m_{\rm ej}$) injection budget of
\begin{align}
    E_{\rm SN} &= 10^{51} \eta_{\rm SN} \frac{m_{*}}{m_{\rm{SN}}}\, \rm{erg}\, ,\\
    m_{\rm ej} &= \eta_{\rm SN} m_{*}\,,
\end{align}
respectively. Here, $\eta_{\rm SN}$ is the fraction of stellar mass that is recycled into SN ejecta; $m_{\rm SN}$ is the average stellar mass of a type II SN progenitor, and $m_{*}$ is the mass of a stellar particle. Assuming a \cite{Chabrier2003} initial mass function, we set $\eta_{\rm SN}=0.2$, and $m_{\rm SN}=20 \rm{M}_{\odot}$.

At each SN explosion, $0.1E_{\rm SN}$ is released in the form of CR energy in the cell hosting the SNR (see Section \ref{CRs} for a description of CR physics). For the energy associated to the thermal component ($0.9E_{\rm SN}$), we used the mechanical feedback prescription detailed in \cite{KimmCen14} and \cite{Kimm15}. In this method, the momentum is deposited into the neighbour cells of a SN hosting cell, with the magnitude adaptively depending on whether the adiabatic phase of the SN remnant is captured by this bunch of cells and the mass within them, or whether the momentum-conserving (snow-plough) phase is expected to have already developed on this scale. If the energy-conserving phase is resolved, the momentum is given by energy conservation, while if it is unresolved, the final momentum, which depends via the cooling efficiency on the density and metallicity, is given by \cite{Blondin1998,Thornton1998}. We injected CR energy into the cell hosting the SN and neighbouring cells using the same relative weights as for the mass and momentum, to determine the CR energy assigned to each neighbour.

We note that CR injection could affect the SNR evolution itself and change the subgrid prescription for mechanical feedback: as a relativistic fluid ($\gamma_{\rm CR}=4/3$), CRs suffer less adiabatic loss than the thermal gas, so that at late times they dominate thermal pressure. Most importantly, CR energy has longer cooling time scales than the thermal energy, and is not radiated away during the snow-plough phase, but rather continues to support SNR expansion \citep{Diesing18}. In future work, we will investigate how the momentum deposition of SNR is affected by the presence of CRs, but that is beyond the scope of this paper.

\subsection{Cosmic ray magnetohydrodynamics}
\label{CRs}

\renewcommand{\arraystretch}{1.2}
\begin{table*}
	\centering
	\caption{Models of CR injection and transport. Runs with anisotropic diffusion also include a perpendicular diffusion coefficient $\kappa_{\perp}=0.01\kappa_{\parallel}$. The names displayed in the first column are used throughout the entire paper.}
	\label{tab2}
	\begin{tabular}{P{3.0cm}P{2cm}P{4.2cm}P{2.9cm}P{2.9cm}}
		\hline
		CR model name & $f_{\rm{cr}}$ & Diffusion & Streaming & Streaming Heating ($-\vec{u}_{\rm{st}} \cdot \nabla P_{\rm CR}$)  \\
		\hline
		noCR & 0 & No & No & No  \\
		Advection & 0.1 & No & No & No  \\
		$\kappa = 3\times 10^{27}\,\rm{cm}^{2}\,\rm{s}^{-1}$ & 0.1 & Anisotropic, $\kappa_{\parallel} = 3\times 10^{27}\,\rm{cm}^{2}\,\rm{s}^{-1}$ & No & No \\
		$\kappa = 1\times 10^{28}\,\rm{cm}^{2}\,\rm{s}^{-1}$& 0.1 & Anisotropic, $\kappa_{\parallel} = 1\times 10^{28}\,\rm{cm}^{2}\,\rm{s}^{-1}$ & No & No\\
		$\kappa = 3\times 10^{28}\,\rm{cm}^{2}\,\rm{s}^{-1}$& 0.1 & Anisotropic, $\kappa_{\parallel} = 3\times 10^{28}\,\rm{cm}^{2}\,\rm{s}^{-1}$ & No & No\\
		$\kappa = 1\times 10^{29}\,\rm{cm}^{2}\,\rm{s}^{-1}$& 0.1 & Anisotropic, $\kappa_{\parallel} = 1\times 10^{29}\,\rm{cm}^{2}\,\rm{s}^{-1}$ & No & No \\
	    Isodiff & 0.1 & Isotropic, $\kappa = 3\times 10^{27}\,\rm{cm}^{2}\,\rm{s}^{-1}$ & No & No\\
	    Streaming & 0.1 & No & Yes, $f_{\rm boost} = 1$ & Yes \\
	    Streaming Boost & 0.1 & No & Yes, $f_{\rm boost} = 4$ & Yes \\
		\hline
	\end{tabular}
\end{table*}

CRs are modelled as a fluid represented by one separate energy equation with one single bin of energy density $e_{\rm CR}$ which contributes to the total energy of the fluid in the equations of ideal MHD:

\begin{align}
\label{1}
    &\frac{\partial \rho}{dt} + \nabla \cdot (\rho \vec{u}) = 0\,,\\
    &\frac{\partial \rho \vec{u}}{dt} +\nabla\cdot \left( \rho \vec{u}\vec{u}+P_{\rm tot} + \frac{\vec{B} \vec{B}}{4 \pi}\right)=\rho \vec{g} +\dot{p}_{\rm SN}\,,\\
    &\frac{\partial e}{dt} +\nabla\cdot \left( (e+P_{\rm tot}) \vec{u} - \frac{\vec{B} (\vec{B}\cdot \vec{u})}{4 \pi}\right) + \nabla \cdot \vec{F}_{\rm st} = \rho \vec{u} \cdot \vec{g} + Q_{\rm CR}\nonumber \\
    &\hspace{2.2cm}+ Q_{\rm th}  - \Lambda_{\rm rad} - \Lambda_{\rm CR} -\nabla\cdot (-\kappa\vec{b}(\vec{b}\cdot \nabla) e_{\rm CR})\,, \\
    & \frac{\partial \vec{B}}{dt} - \nabla \times (\vec{u}\times\vec{B}) = \vec{0}\,, \\
    &\frac{\partial e_{\rm CR}}{dt} +\nabla\cdot( e_{\rm CR} \vec{u}) + \nabla \cdot \vec{F}_{\rm st} = -P_{\rm CR}\nabla\cdot \vec{u} + Q_{\rm CR} - \Lambda_{\rm CR} \nonumber \\ & \hspace{3cm} -\nabla\cdot (-\kappa\vec{b}(\vec{b}\cdot \nabla) e_{\rm CR}) -\vec{u}_{\rm{st}} \cdot \nabla P_{\rm CR}\, , 
\end{align}
where $\rho$ is the density, $\vec{u}$ is the gas vertical velocity, and $\vec{B}$ the magnetic field. The total energy density includes the kinetic, internal, CR and magnetic energy densities, that is $e= 0.5 \rho \vec{v}^2 + e_{\rm th} + e_{\rm CR} + \vec{B}^2/8\pi$, where $e_{\rm th}$ is the internal energy; the total pressure includes the magnetic, thermal ($P_{\rm th}$) and CR ($P_{\rm CR}$) pressures, that is $P_{\rm tot}=  P_{\rm th} + P_{\rm CR} + \vec{B}^2/8\pi$, where $P_{\rm CR}$ and $P_{\rm th}$ are related to their energy density with $e_{\rm th}=  P_{\rm th}/(\gamma-1)$ and $e_{\rm CR}=  P_{\rm CR}/(\gamma_{\rm CR}-1)$, with $\gamma=5/3$ and $\gamma_{\rm CR}=4/3$ for a purely monoatomic ideal thermal component and a fully relativistic population of CRs; $\vec{g}$ is the gravitational field; $\dot{p}_{\rm SN}$ is the source therm for momentum injection by SNe; $Q_{\rm th}$ is the thermal energy source term and includes UV background heating as well as CR collisional heating due to Coulomb collisions \citep{GuoOh2008}; $Q_{\rm CR}$ is the energy source term for CRs from SNe explosions;  $\Lambda_{\rm rad}$ is the radiative cooling of the thermal component; $\Lambda_{\rm CR}$ is the total CR energy loss rate due to Coulomb and hadronic collisions \citep{GuoOh2008}: $\Lambda_{\rm CR}=7.51\times 10^{-16}(n_{\rm e}/{\rm cm^{-3}})(e_{\rm CR}/{\rm erg\,cm^{-3}}) \,\rm erg\,s^{-1}\,cm^{-3}$, with a corresponding reinjection to the thermal component of $2.63\times 10^{-16} (n_{\rm e}/{\rm cm^{-3}})(e_{\rm CR}/{\rm erg\,cm^{-3}})\,\rm erg\,s^{-1}\,cm^{-3}$ included in $Q_{\rm th}$.
CR anisotropic diffusion term $-\nabla \cdot (-\kappa \vec{b}(\vec{b}. \nabla)e_{\rm CR})$ (where $\vec{b}$ is the magnetic field unit vector) is solved with the implicit solver of~\cite{DuboisCommercon16}.
The value of the constant diffusion coefficient $\kappa$ is varied within the range $[3\times 10^{27},10^{29}]\, \rm cm^2\,s^{-1}$ (see Table \ref{tab2}).
Typical inferred values of the (isotropic) diffusion coefficient in the Milky Way from observational constraint obtained with {\sc galprop} lie between $3-8\times 10^{28}\, \rm cm^2\, s^{-1}$ for particles at a few GeV~\citep{Strong07,Trotta11}. Hydrodynamical simulations with a fluid treatment of the CR physics produce consistent scaling of the gamma-ray luminosity from CR hadronic losses with SFR~\citep{Ackermann12} or gas surface density~\citep{Lacki11} using a broad range of diffusion coefficient $10^{28}-3\times 10^{29}\rm \, cm^2\,s^{-1}$~\citep{SalemBryan16, Pfrommer17gray, Chan2018}.

As CRs interact with Alfv\'en waves, they stream along their own gradient of pressure.
This streaming of CRs is represented by the advection term $\vec{F}_{\rm st}=f_{\rm{b}}\vec{u}_{\rm st}(e_{\rm CR}+P_{\rm CR})$ and the associated cooling (for CRs) and heating (for the thermal component) term $-\vec{u}_{\rm{st}}.\nabla P_{\rm CR}$, where $\vec{u}_{\rm st} =  {\rm sign}(\vec{b} . \nabla e_{\rm CR}) \vec{u}_{\rm A}$ is the CR streaming velocity ($\vec{u}_{\rm A}=\vec{B}/\sqrt{4\pi\rho}$ is the Alfv\'en speed, thus, the streaming is also fully anisotropic along the magnetic field lines).
$f_{\rm b}$ is the boost factor of the streaming velocity and accounts for the fact that Alfv\'en waves might experience damping by various mechanisms, allowing for weaker confinement by streaming, and, hence, a faster streaming velocity.

The ISM includes a cold, dense phase, where the effect of ion-neutral
  damping should be accounted for. The imperfect ionisation of the thermal particles causes CRs to couple less
  efficiently to the gas in this cold phase and allows them to escape
  these regions at velocities faster than the Alfv\'en velocity, while at the same time limiting the local diffusion coefficient~\citep{Wiener2013,Nava16,Brahimi19}. As CRs enter the completely ionised warm and hot phases of the ISM, they establish a tight coupling while the weaker non-linear Landau damping
  dominates the wave damping~\citep{Wiener2013,Wiener2017,Thomas19}.
  Turbulent damping of waves, effective in the various phases of the ISM, can control both the amount of coupling of CRs with the plasma and the level of anisotropy of the diffusive flux depending on the nature of the magnetised turbulence~\citep{Yan2002,Farmer04, Lazarian16,Holguin18,Nava19}. 
Finally, the shape of the spectrum of CRs, harder (flatter) close to acceleration sites (i.e. SN remnants), is expected to play an important role in how well CRs are tight to the plasma by the spectrum-dependent excited waves.
  In order to simplify the complex mechanisms of wave damping that change with the different ISM phases, in this work: i) we have assumed diffusion with an effective galactic-wide constant coefficient (and which degree of anisotropy does not vary either), ii) we have assumed streaming with a uniform value of velocity, and iii) we have tested the effect of diffusion and streaming separately (while in practice their are intertwined and spatially varying). 
  In this `effective' approach, the streaming velocity entering the advection term is boosted by a uniform factor of $f_{\rm b}=4$ (not the heating term of streaming), which is a representative value for CR-excited waves that are damped mostly by turbulence in the diffuse ISM~\citep[see][for details]{Ruszkowski17}. 
  We keep a more in-depth numerical investigation of wave-coupling effects for future work (Brahimi et al. in prep.).

While the heating term of streaming is straightforward to implement with a finite-differentiation, the advection term of streaming, if solved with a standard explicit Godunov-like approach puts stringent constraints on the time step $\Delta t\propto\Delta x^3$ due to the non-continuous nature of the streaming velocity at energy extrema \citep[see][for discussion]{Sharmaetal09}.
A manipulation of the different terms in the streaming flux allows one to express this term equivalently as a pure diffusion term, which can then be added to the constant diffusion term $\kappa$, and to use the implicit solver for anisotropic diffusion and streaming~\citep[see][for details and tests of the method]{Dubois19}.

For our fiducial galaxy G9, we ran nine different models: a run without CRs ("noCR"); a run with CR injection but no transport ("Advection"); a run with isotropic diffusion of CRs ("Isodiff"); 4 runs with anisotropic diffusion where we varied the value of the diffusion coefficient $\kappa$ parallel to the magnetic field; two runs with streaming where we used a boost factor of either $f_{\rm boost} = 1$ ("Streaming") or1 $f_{\rm boost} = 4$ ("Streaming boost"). 
We present in Table~\ref{tab2} the different CR models presented in this paper. The Isodiff model has been run with a diffusion coefficient of  $\kappa = 3\times 10^{27}\,\rm{cm}^{2}\,\rm{s}^{-1}$: since in that model, CR diffusion is identical in all directions, the amount of CR diffusion is expected to be theoretically equivalent to the anisotropic diffusion run with $\kappa = 1\times 10^{28}\,\rm{cm}^{2}\,\rm{s}^{-1}$.

The original implementation of the implicit diffusion solver introduced in~\cite{DuboisCommercon16} has been here extended to include a minmod slope limiter to the transverse gradient of the CR energy density for the centred asymmetric scheme as proposed by~\cite{Sharma07} to preserve the monotonicity of the flux. 
Since the limiter produces a non-symmetric matrix, the linear system is solved with a biconjugate gradient stabilised method. This procedure guarantees that the values of the CR energy density obtained by the anisotropic diffusion/streaming implicit solver are always positives\footnote{All the simulations have also been run with the original unlimited version, which does not guarantee the positivity of the solution. Thus, in that case, the CR energy can become negative, and has to be readjusted. This translates into a spurious CR energy injection upon SN explosions, that depends on the diffusion coefficient: namely at each SN explosion, an additional 3\%, 7\%, 12\%, 15\% of CR energy (relative to the expected CR energy to be injected) is injected for diffusion coefficients of respectively $3 \times 10^{27}\,\rm{cm}^{2}\rm{s}^{-1}$, $1 \times 10^{28}\,\rm{cm}^{2}\rm{s}^{-1}$, $3 \times 10^{28}\,\rm{cm}^{2}\rm{s}^{-1}$, $1 \times 10^{29}\,\rm{cm}^{2}\rm{s}^{-1}$. The results in terms of, for instance, mass outflow rates or star-formation rates do not differ by more than $\sim 10\%$ with the current positivity-preserving version of the solver.}.

\section{Case study of the G9 galaxy}
\label{results}

We begin by comparing the different models of CR injection and transport on the wind and ISM morphology, star formation, outflow generation, and CR energy distribution of the G9 galaxy.

\begin{figure*}
   \centering
   \subfloat[Density (cm$^{-3}$)]{\includegraphics[width=0.5\columnwidth]{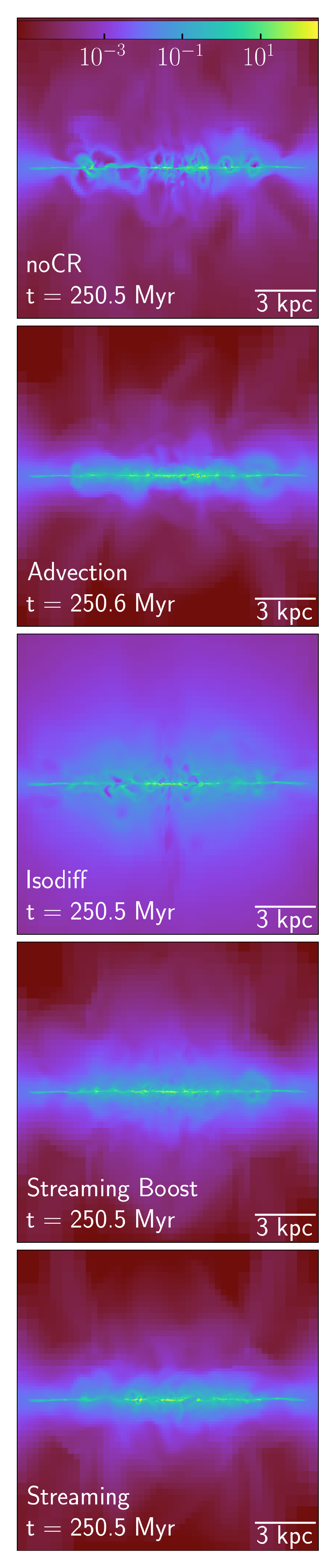}}
   \subfloat[Vertical velocity ($\rm km\,s^{-1}$)]{\includegraphics[width=0.5\columnwidth]{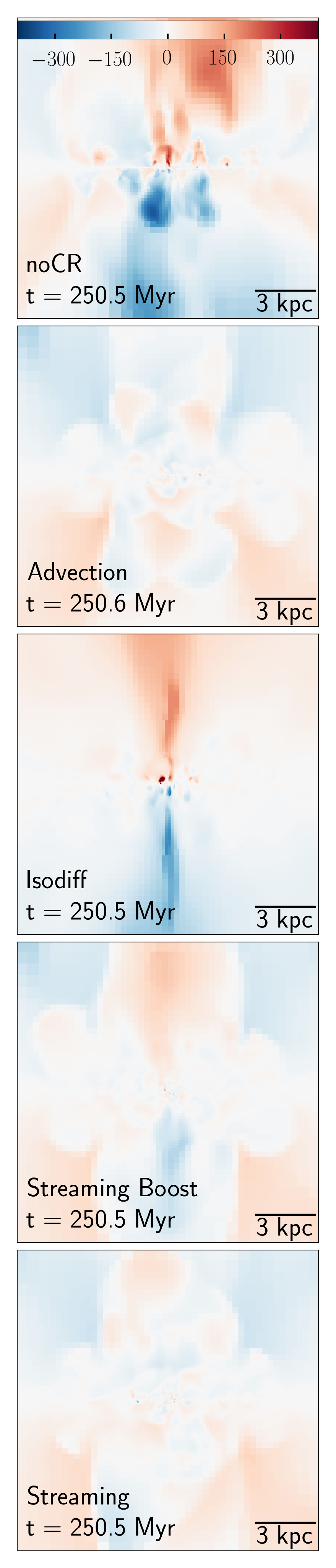}}
\subfloat[$P_{\rm CR}+P_{\rm th}$ ($\rm erg\,cm^{-3}$) ]{\includegraphics[width=0.5\columnwidth]{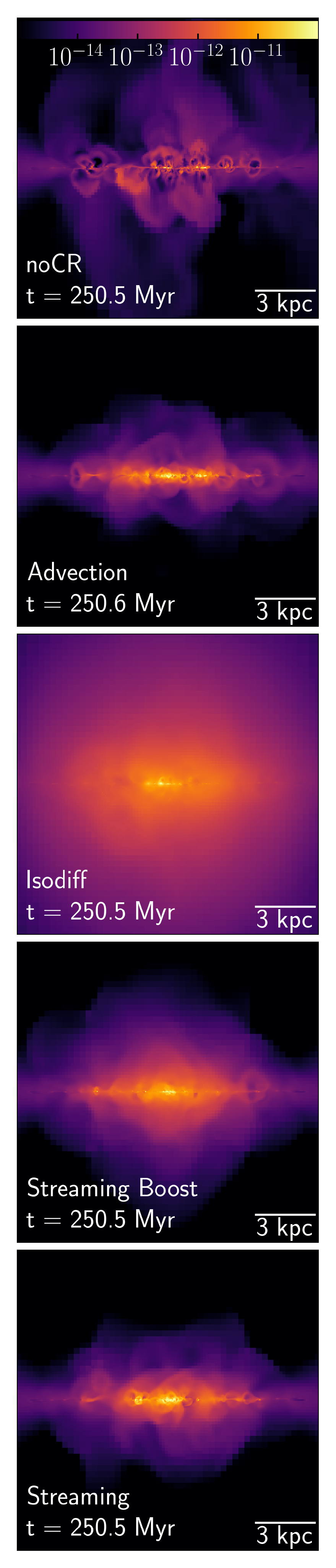}}
   \subfloat[$P_{\rm CR}/P_{\rm th}$ ]{\includegraphics[width=0.5\columnwidth]{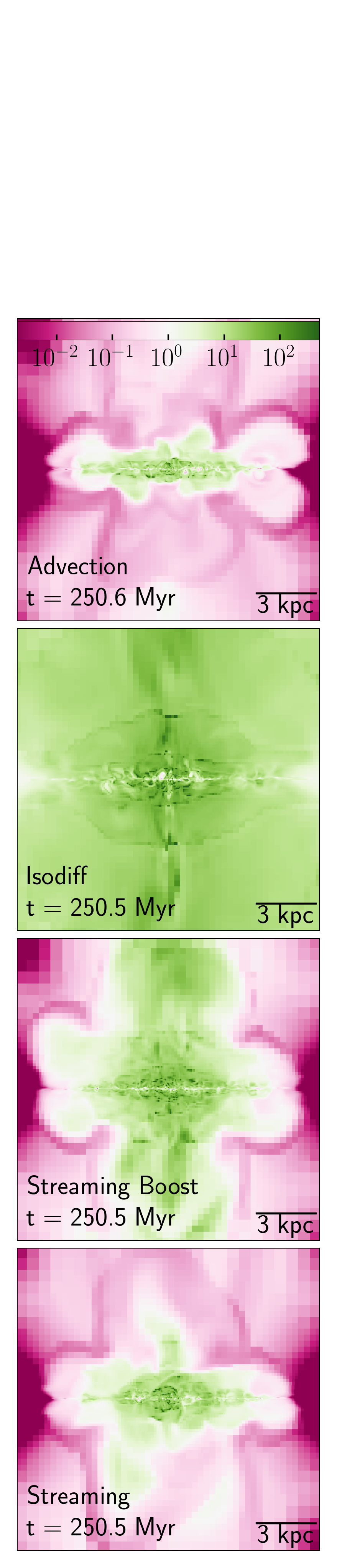}}
 \caption{ Slices of the G9 galaxy, from left to right: gas density, gas vertical velocity, sum of thermal and CR pressures, and ratio of CR to thermal pressure, seen edge-on at 250 Myr for the different simulations as indicated in the panels (one model per row). See Fig.~\ref{edgeon2} for the four remaining runs. In the noCR case, a low density wind is generated with velocities of a few hundred $\rm{km}\,{s}^{-1}$; in the Advection model, the disc is puffed up but the wind is much weaker. When adding CR diffusion, the wind is 10 times denser with velocities similar to noCR case. The sum of CR and thermal pressures, $P_{\rm CR}+P_{\rm th}$, increases by 2 orders of magnitude and mainly consists of CR pressure. With CR streaming, the morphology is similar to the Advection case, but with a mild wind.}
    \label{edgeon1}
 \end{figure*}
 
 \begin{figure*}
   \centering
   \subfloat[Density (cm$^{-3}$)]{\includegraphics[width=0.5\columnwidth]{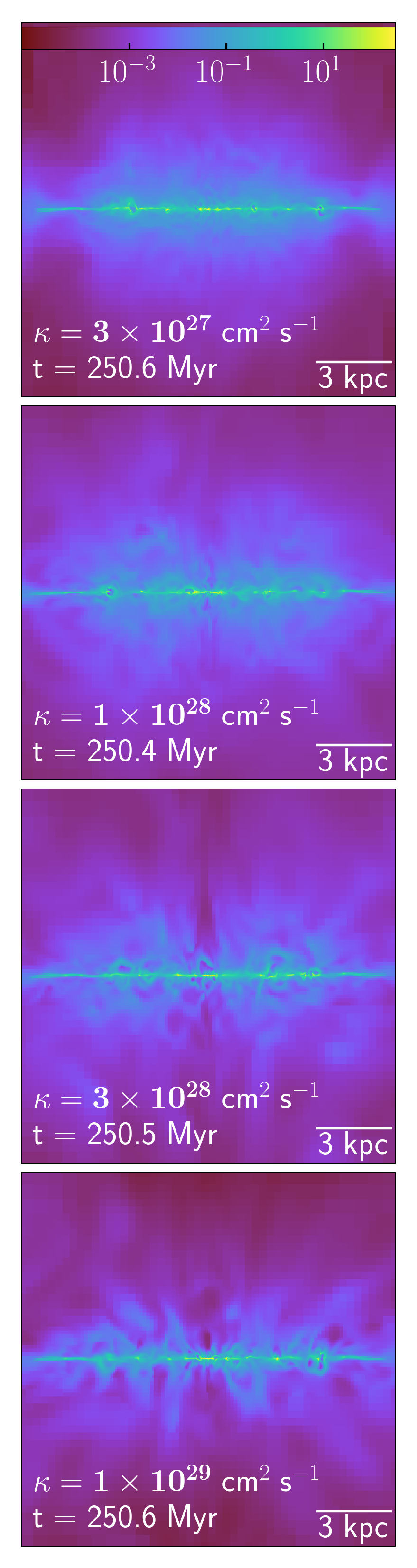}}
   \subfloat[Vertical velocity ($\rm km\,s^{-1}$)]{\includegraphics[width=0.5\columnwidth]{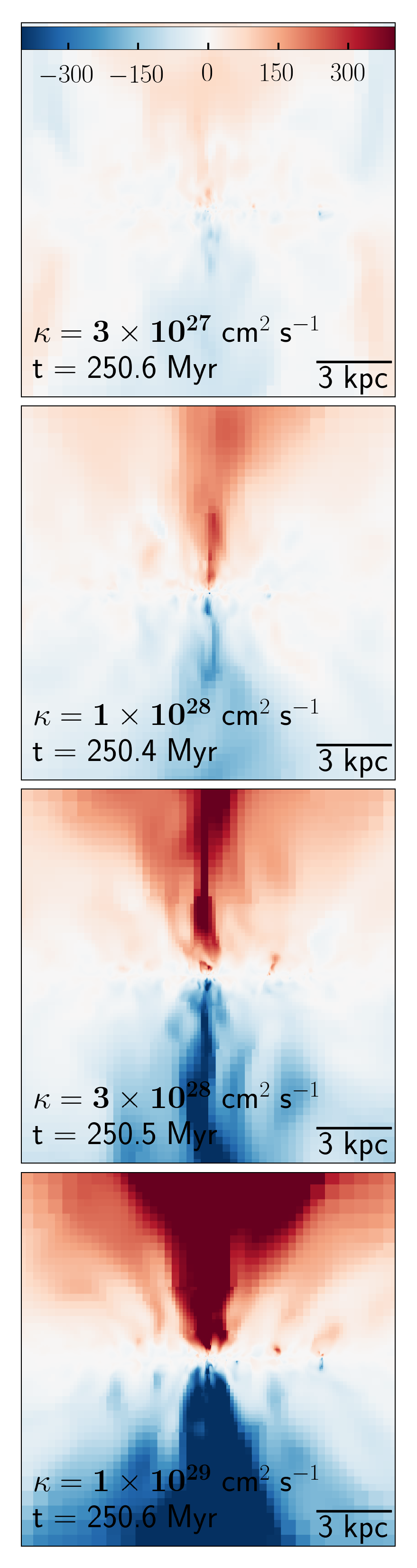}}
\subfloat[$P_{\rm CR}+P_{\rm th}$ ($\rm erg\,cm^{-3}$) ]{\includegraphics[width=0.5\columnwidth]{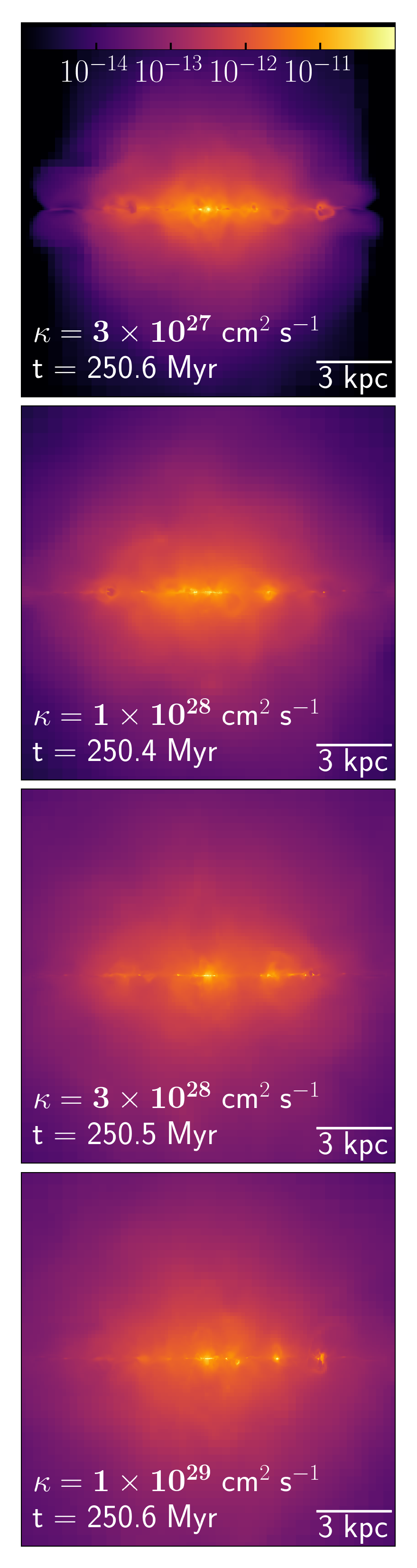}}
   \subfloat[$P_{\rm CR}/P_{\rm th}$ ]{\includegraphics[width=0.5\columnwidth]{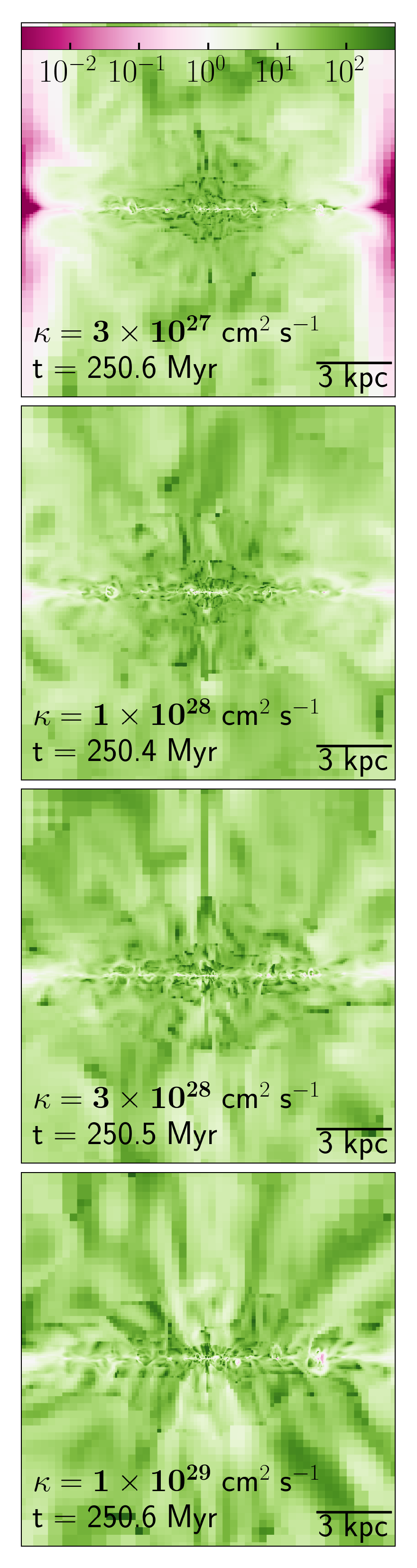}}
 \caption{Slices of the G9 galaxy, from left to right: gas density, gas velocity, sum of CR and thermal pressures, and ratio of CR to thermal pressure, seen edge-on at 250 Myr for the different simulations as indicated in the panels (one model per row). See Fig.~\ref{edgeon1} for the other runs. In the noCR case, a low density wind is generated with velocities of a few hundred $\rm{km}\,{s}^{-1}$; in the Advection model, the disc is puffed up but the wind is much weaker. When adding CR diffusion, the wind is 10 times denser with velocities similar to noCR case. The sum of CR and thermal pressures increases by 2 orders of magnitude and mainly consists of CR pressure. With CR streaming, the morphology is similar to the Advection case, but with a mild wind.}
    \label{edgeon2}
 \end{figure*}
 
 \begin{figure*}
   \centering
   \subfloat[Density (cm$^{-3}$)]{\includegraphics[width=0.5\columnwidth]{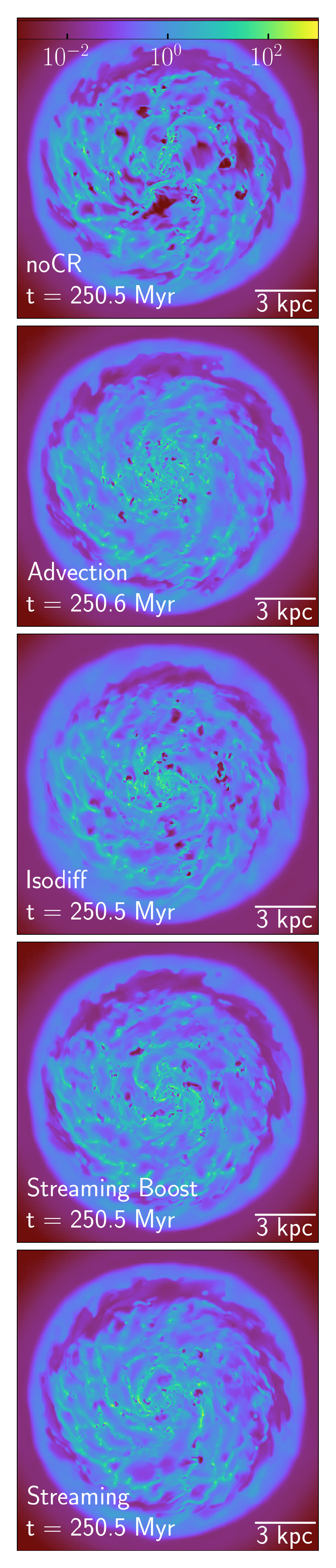}}
   \subfloat[][Density (zoomed-in, cm$^{-3}$)]{\includegraphics[width=0.5\columnwidth]{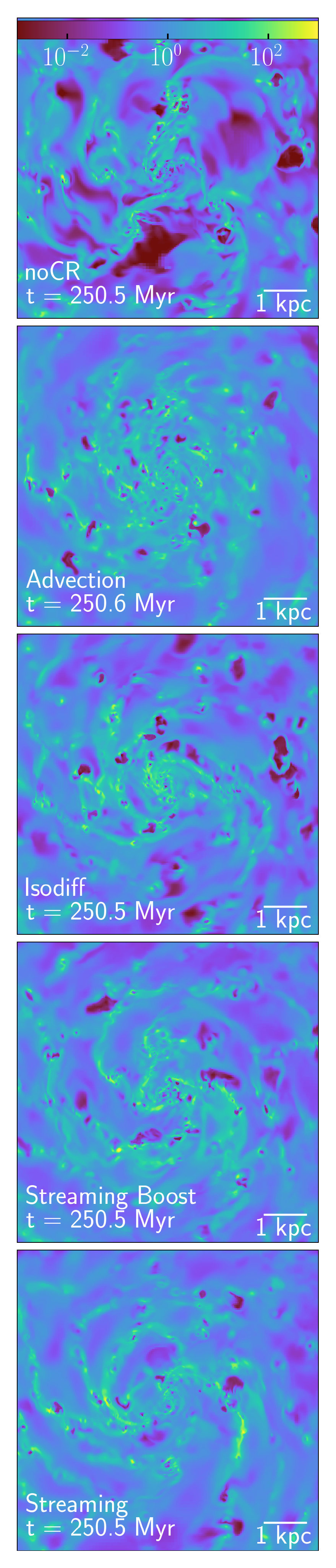}}
   \subfloat[$P_{\rm CR}+P_{\rm th}$ ($\rm erg\,cm^{-3}$) ]{\includegraphics[width=0.5\columnwidth]{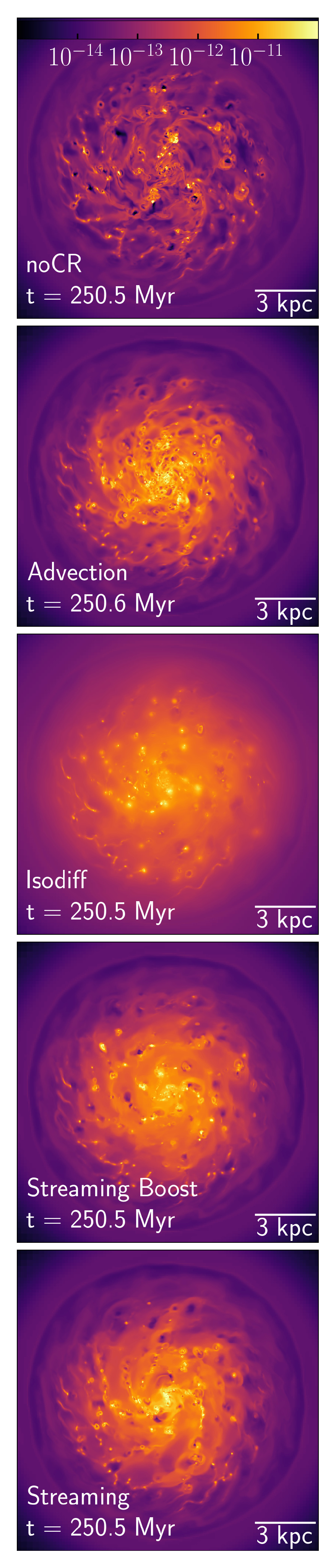}}
   \subfloat[$P_{\rm CR}/P_{\rm th}$ ]{\includegraphics[width=0.5\columnwidth]{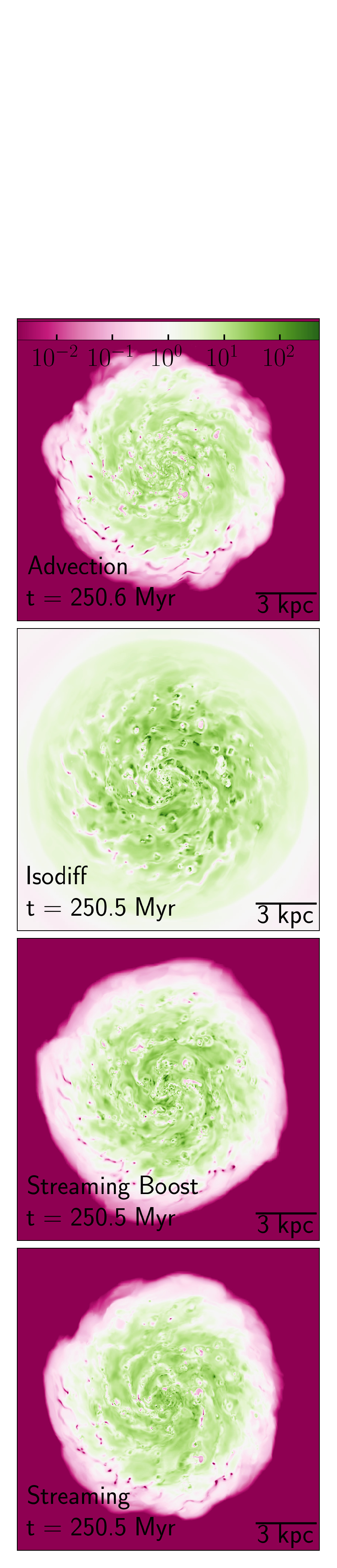}}
  \caption{Slices of the G9 galaxy, from left to right: gas density, slices of gas density zoomed-in compared to the left hand panels, sum of CR and thermal pressures, and ratio of CR to thermal pressure, seen edge-on at 250 Myr for the different simulations as indicated in the panels (one model per row). See Fig.~\ref{faceon2} for the four remaining runs. The density distributions in the disc is smoother with CR injection. The sum of CR and thermal pressures in the disc increases by 1 -- 2 orders of magnitude when adding CRs, and is either dominated by CR pressure, or equally distributed between CR and thermal pressure.}
   \label{faceon1}
 \end{figure*}
 
  \begin{figure*}
   \centering
   \subfloat[Density (cm$^{-3}$)]{\includegraphics[width=0.5\columnwidth]{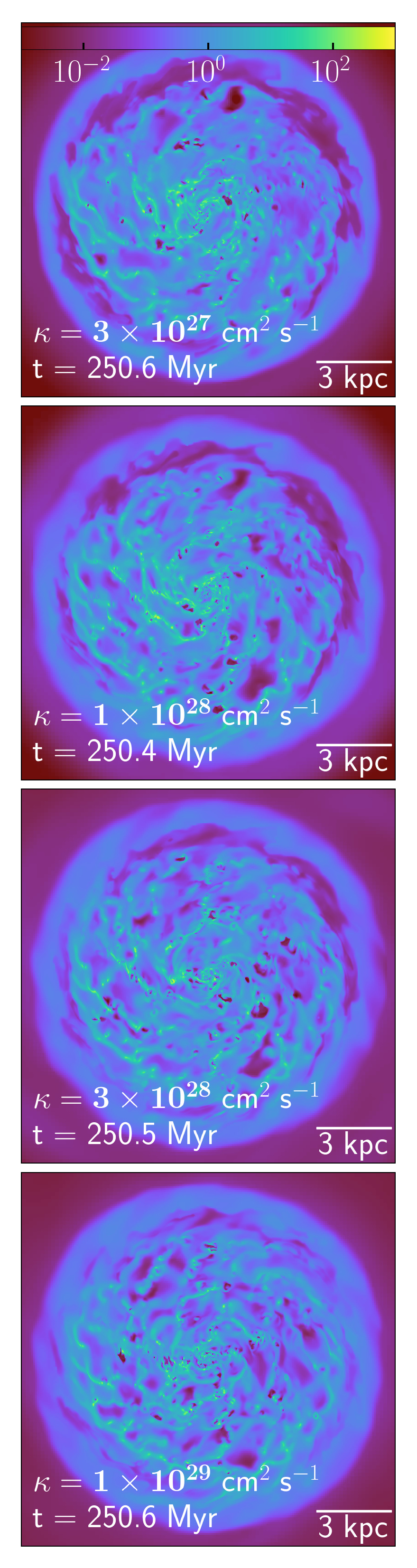}}
   \subfloat[][Density (zoomed-in, cm$^{-3}$)]{\includegraphics[width=0.5\columnwidth]{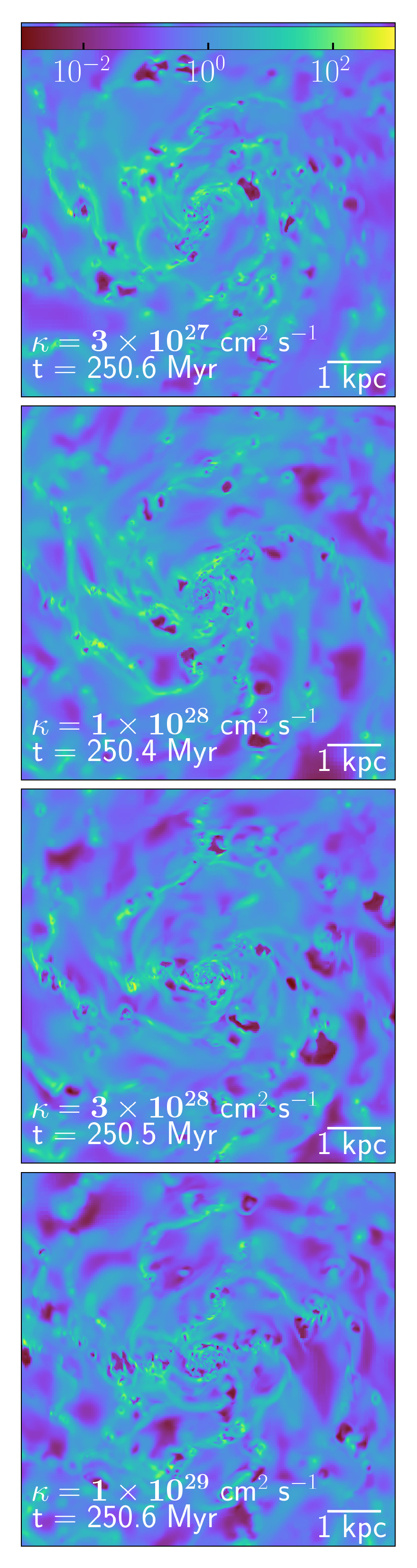}}
   \subfloat[$P_{\rm CR}+P_{\rm th}$ ($\rm erg\,cm^{-3}$) ]{\includegraphics[width=0.5\columnwidth]{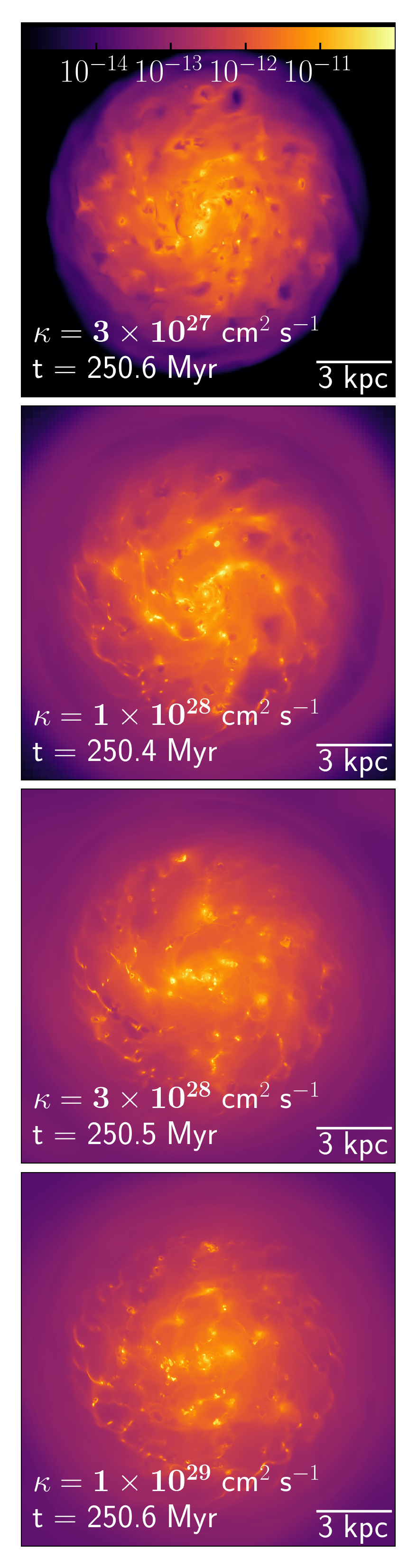}}
   \subfloat[$P_{\rm CR}/P_{\rm th}$ ]{\includegraphics[width=0.5\columnwidth]{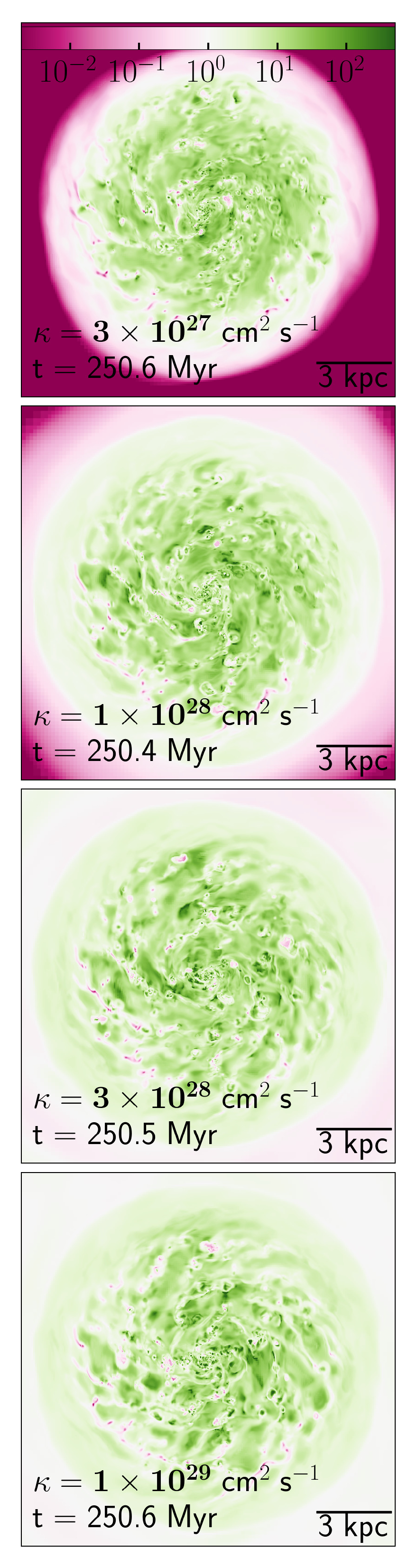}}
  \caption{Slices of the G9 galaxy, from left to right: gas density, slices of gas density zoomed-in compared to the left hand panels, sum of CR and thermal pressures, and ratio of CR to thermal pressure, seen edge-on at 250 Myr for the different simulations as indicated in the panels (one model per row). See Fig.~\ref{faceon1} for the other runs. The density distributions in the disc is smoother with CR injection. The sum of CR and thermal pressures in the disc increases by 1 -- 2 orders of magnitude when adding CRs, and is either dominated by CR pressure, or equally distributed between CR and thermal pressure.}
   \label{faceon2}
 \end{figure*}
 
\subsection{Galaxy and wind morphologies}

Figures~\ref{edgeon1} and~\ref{edgeon2} show the density, the velocity, the sum of CR and thermal pressures and ratio of CR to thermal pressures in 15 kpc-wide slices of the G9 galaxy viewed edge-on. Without CR injection (noCR), a low density wind ($\leq 10^{-3}\, \rm{cm}^{-3}$) is generated with velocities of a few hundred $\rm km\,s^{-1}$. With the injection of CR energy without any transport (Advection), the disc is puffed up but the wind is much weaker. CR streaming allows for a mild wind formation, weaker than in the simulations with CR diffusion. When adding isotropic diffusion of CRs, the wind is 10 -- 100 times denser with similar velocities to the noCR case. The sum of CR and thermal pressures at 1--5 kpc from the mid-plane increases by 2 orders of magnitude and mainly consists of CR pressure. This strong effect on wind density and pressure is slightly mitigated but still substantial when modelling anisotropic diffusion. As already pointed out, the amount of CR diffusion in the Isodiff run is roughly equal to that of the anisotropic run with $\kappa = 1\times 10^{28}\,\rm{cm}^{2}\,\rm{s}^{-1}$, but one sees that the wind is weaker in the $\kappa = 1\times 10^{28}\,\rm{cm}^{2}\,\rm{s}^{-1}$ compared to the Isodiff run, which suggests that the azimuthal component of the magnetic field is stronger than its vertical component. We confirm this later in Section~\ref{section:bfieldevol}. The velocity of the winds increases with the anisotropic diffusion coefficient, and the density in the first 5 kpc away from the disc plane decreases at higher diffusion coefficients.
The Streaming run do not differ much from the Advection case, only when streaming is boosted by a factor of 4, do we see a larger wind density and velocity.

Figures~\ref{faceon1} and~\ref{faceon2} show the differences between the different models for the G9 galaxy viewed face-on. Unlike in the edge-on view, the density distributions in the ISM are more similar visually, but smoother with CR injection, as opposed to the presence of large pockets of low density gas in the noCR simulation. The injection of CR smooths the density distribution (as it will be seen later more quantitatively). This effect is mitigated at higher diffusion coefficient, because CR energy diffuses out more quickly. The sum of CR and thermal pressures in the disc increases by 1 -- 2 orders of magnitude when adding CRs, and is either dominated by CR pressure by large in the diffuse parts of the ISM -- and increasingly with the increase of the value of the diffusion coefficient --, or equally distributed between the CR and the thermal pressure in the cold filamentary ISM. 

\subsection{Density distribution and star formation}

\begin{figure}
    \centering
    \includegraphics[width=\columnwidth]{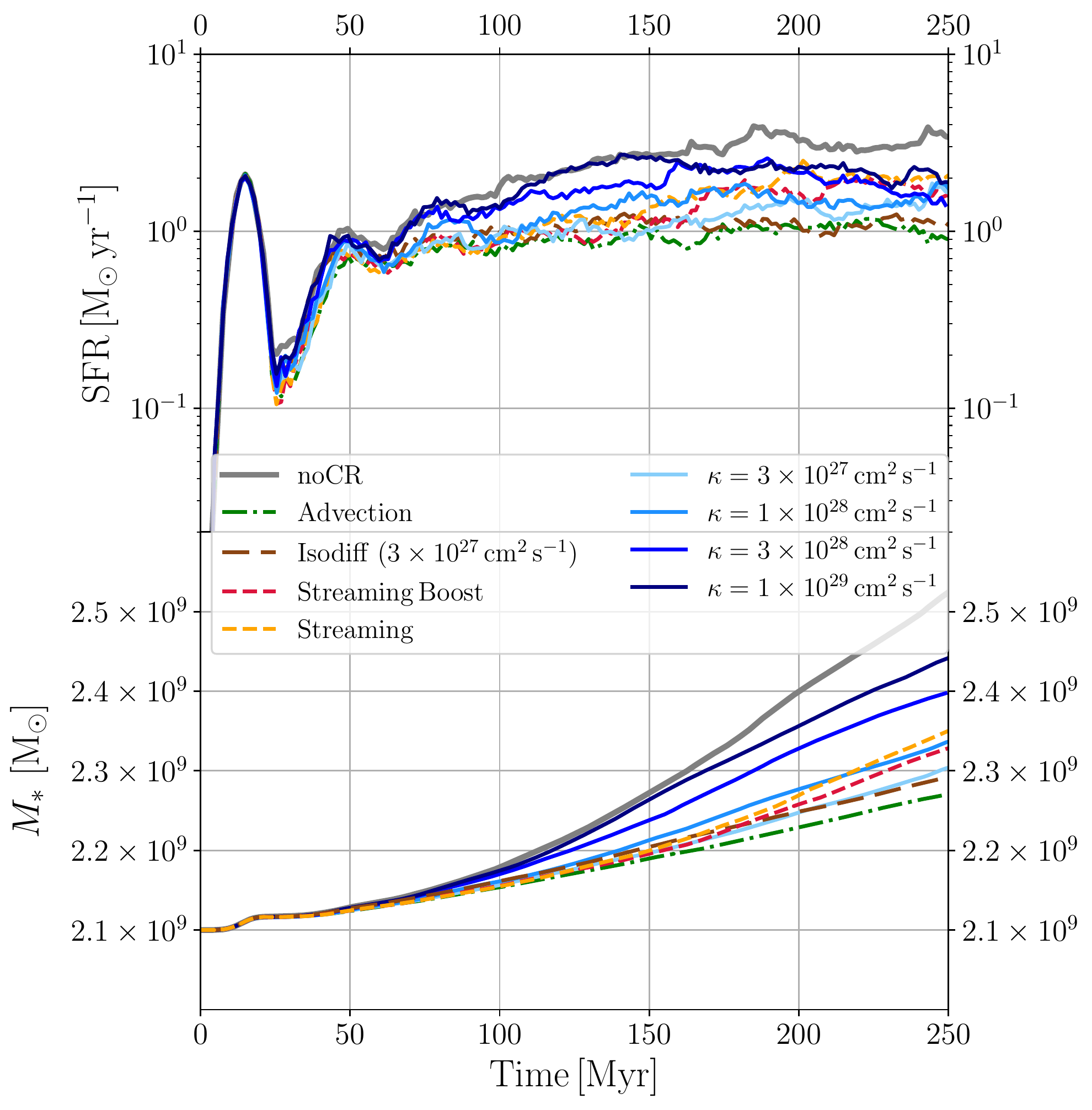}
    \caption{\emph{Top panel:} SFR as a function of time for the G9 galaxy with and without CR injection, with different CR transport models. \emph{Bottom panel:} Stellar mass as a function of time. Without CR injection (grey thick line), the SFR is 2 -- 3 times greater than with CR injection. The simulation with advection of the CRs is where the SFR is the most affected because CR energy is trapped in the regions of star formation and does not escape.}
    \label{sfr}
\end{figure}

\begin{figure}
  \centering
  \includegraphics[width=0.8\columnwidth]{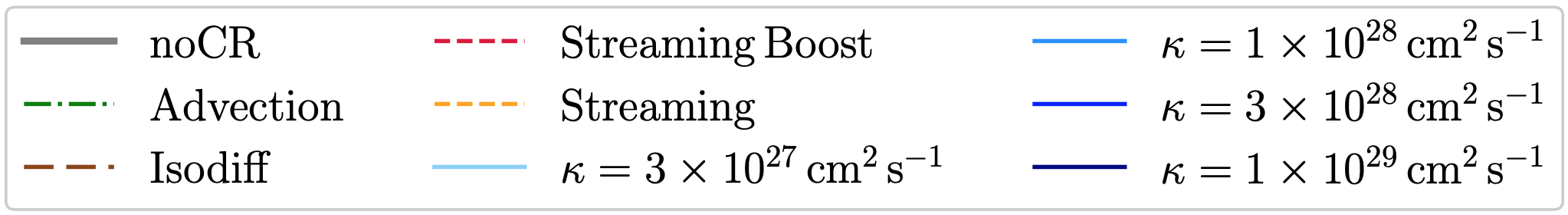}\vspace{-0.3cm}
  \includegraphics[width=0.85\columnwidth]{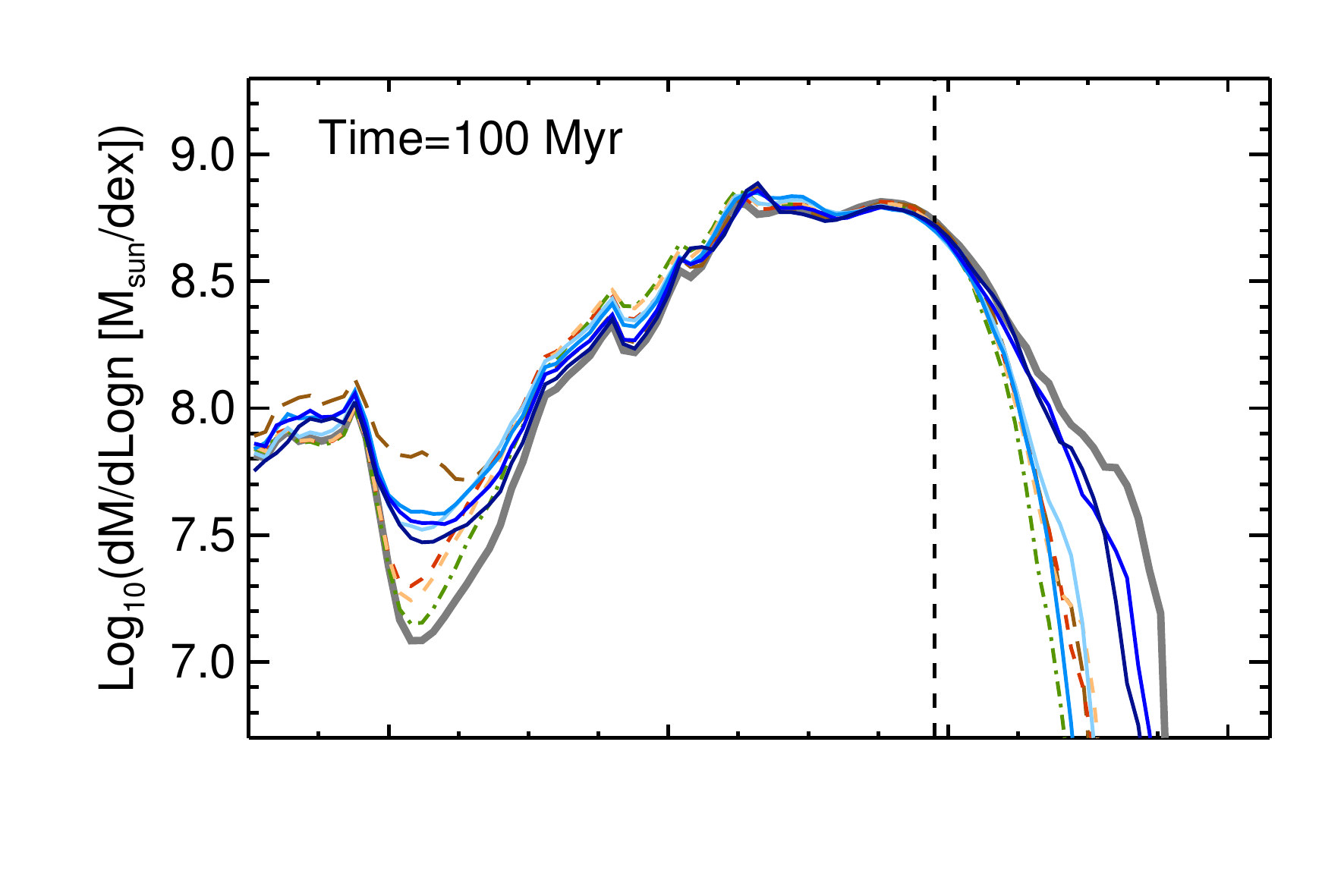}\vspace{-1.37cm}
  \includegraphics[width=0.85\columnwidth]{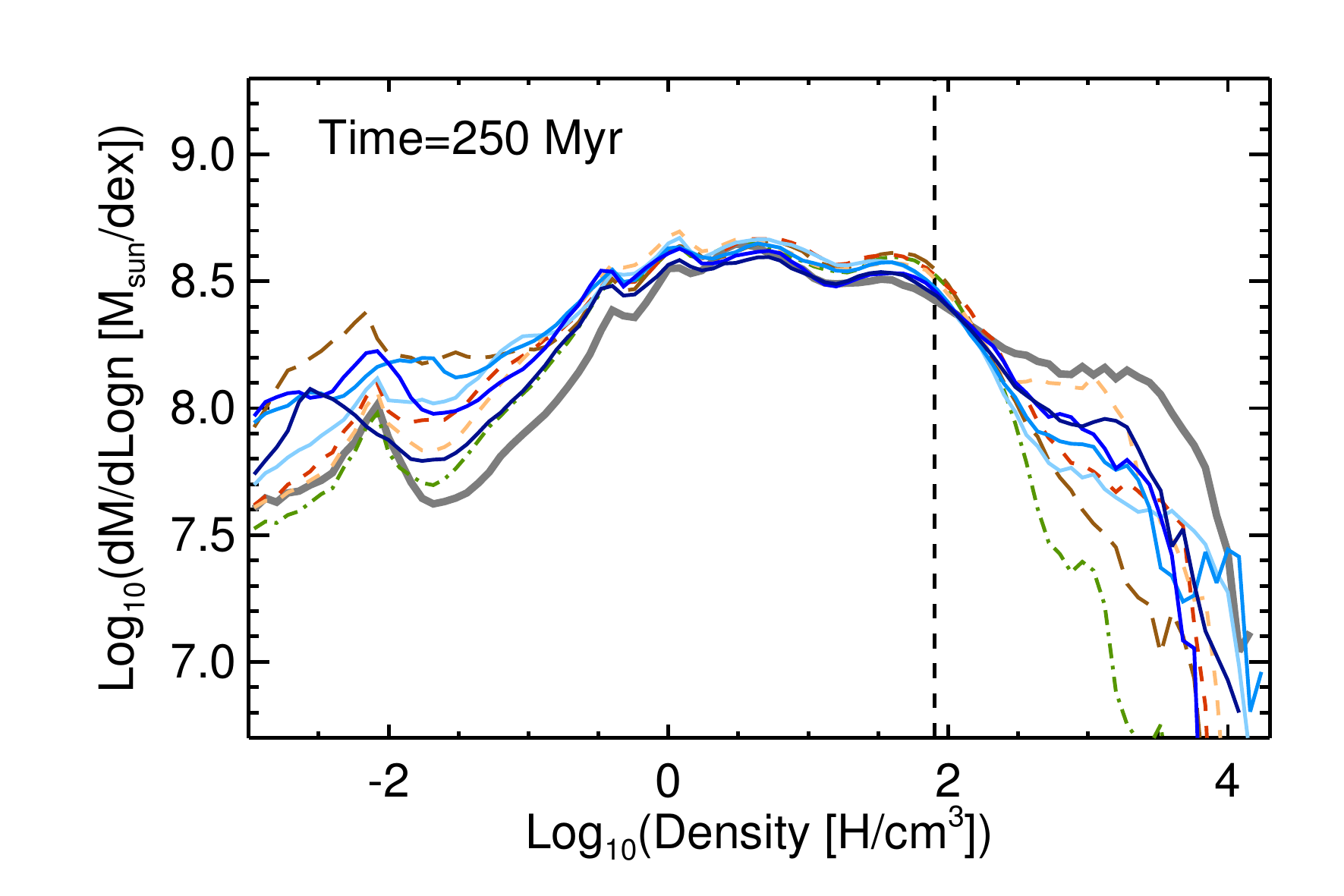}
  \caption{Mass-weighted probability density function of the gas density within a disc of 4 kpc height and 10 kpc radius, centered on the G9 galaxy, for the different runs, at $t=100\,\rm Myr$ (top panel) and $t=250\, \rm Myr$ (bottom panel) throughout the simulations as indicated. The vertical black dashed line shows the density threshold for star formation. Because it adds an additional pressure support against gravity that cools less efficiently than thermal pressure, CR injection reduces the high-density tail of the probability density function and increases the low-density tail.}
  \label{densitypdf}
\end{figure}

Figure~\ref{sfr} shows the SFR and stellar mass of the G9 galaxy as a function of time. Adding CR injection from SNe reduces the SFR by a factor of 2--3 after 250 Myr. The simulation without any CR transport (Advection) is the most efficient at reducing the SFR because the CR pressure is only advected with the gas and therefore remains in the star-forming regions where it smooths out the highest densities (see Fig.~\ref{densitypdf}) and regulates gas infall in dense clouds. In accordance with recent work from \cite{Chan2018} and \cite{Hopkins2019}, we find that overall, CR injection has a modest effect on star formation. The simulations with the higher diffusion coefficient is where the SFR is the least affected because, as already pointed out, CR energy escapes from the star-forming regions and the pressure support against gravitational collapse in the ISM is weaker than for lower diffusion coefficients (see Section \ref{sub:CRenergy}).

Figure~\ref{densitypdf} shows the instantaneous mass-weighted probability density function (PDF) of the density in the ISM (typical wind densities are below $n<10^{-2}\,\rm H\,cm^{-3}$ as it will be shown in Section~\ref{section:mag}) within a disc of 10 kpc radius and 4 kpc height at $t=100$ and $t=250\,\rm Myr$, for the nine different models\footnote{Although it is a small correction, all PDF are renormalised to that of the noCR run, so that their integrated value leads to the same amount of total mass.}.
The injection of CR energy systematically reduces the fraction of gas above the density threshold for star formation, which explains the decrease of the SFR observed in Fig~\ref{sfr}. The Advection model is the one where that fraction is the most suppressed, the reason being that the CR-pressure support in the star-forming regions is higher because the injected energy does not escape: we will see in Section \ref{sub:CRenergy} that the confinement of CRs in higher density regions also leads to more CR-energy loss since CR cooling is higher at higher densities. At early times ($t=100\,\rm Myr$), the larger the diffusion coefficient is, the closer is the density PDF -- and, hence, the SFR -- to the run without CRs, which is reminiscent of a more uniform CR pressure in the disc, and Streaming runs are close to the result without any transport. Though, here, CR radiative losses and energy-injection by SNe complexifies the whole picture, the qualitative behaviour of the density PDF is consistent with simulations of turbulence in the cold neutral medium with CRs from~\cite{Commercon19}: when CR diffusion is large enough, CR pressure cannot be trapped in high gas density regions and the density PDF becomes similar to that without CRs.
At late times ($t=250 \, \rm Myr$), the picture is more complex between the different diffusion coefficients but the qualitative behaviour as opposed to the Advection run is similar to that at early times, and the Streaming runs are now comparable to the runs with anisotropic diffusion.
Conversely, CR injection increases the amount of gas at lower densities, and this effect is larger with diffusion, and CR streaming is the transport model for which this effect is the weakest at early times ($t=100 \, \rm Myr$) and becomes comparable to models with large diffusion coefficient at late times ($t=250 \, \rm Myr$). 

\subsection{Outflows and mass loading}
\label{section:mag}

\begin{figure}
  \centering
  \includegraphics[width=\columnwidth]{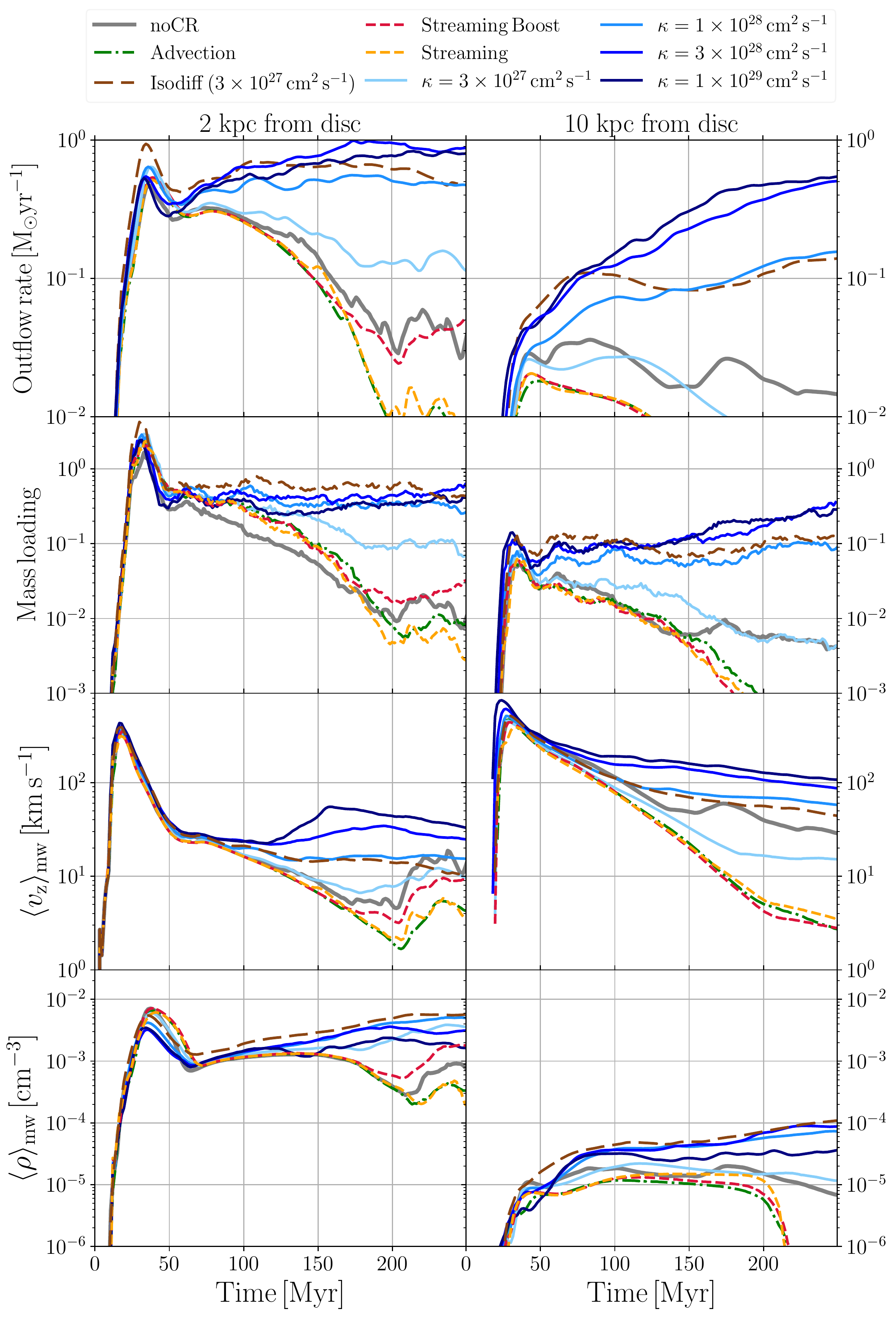}
  \caption{Mass outflow rates of the outflowing gas, mass loading factors, mass-weighted average velocities and mass-weighted average density of the outflowing gas across planes at 2 (left column) and 10 kpc (right column) above the disc plane of the G9 galaxy. The outflow rate quickly drops without CR injection and transport, in the noCR and Advection models. In the Advection-only model, the lack of CR transport prevents the generation of sustained outflows. The outflow rate in the Streaming Boost simulation is very similar to the Advection case at early times and rises to higher values at later times, but the outflow rate is sill weaker than with the other transport models. The isotropic diffusion model and anisotropic diffusion with $\kappa= 3 \times 10^{28} - 1\times 10^{29}\, \rm{cm}^{2} \rm{s}^{-1}$ are more than 10 times more efficient at driving winds compared to the noCR case. The mass average outflow velocity is less affected by the injection of CRs: it is increased at most by a factor of 2 for the highest diffusion coefficient, and is even lower for $\kappa = 3 \times 10^{27}\,\rm{cm}^{2}\rm{s}^{-1}$.}
  \label{outflows}
\end{figure}

In Fig.~\ref{outflows}, we compare the time-evolution of outflows from the G9 galaxy in the different models. We measure the galactic-wide outflow properties across planes parallel to the galaxy disc, at a distance of 2 kpc in the left-hand panels and further out at 10 kpc in the right-hand panels. We show the mass outflow rate, the mass loading factor (the mass outflow rate divided by the SFR), the mass-weighted average velocity and the mass-weighted average density. To measure the mass outflow rate, we select the cells that cross the plane across which we compute the outflow rate, and sum the outflow rates $\dot{m}_{\rm cell}$, obtained in individual cells as: $\dot{m}_{\rm cell}=\rho_{\rm cell} u_{\rm z,cell} \Delta_{\rm x,cell}^2$, where $\rho_{\rm cell}$, $u_{\rm z,cell}$ and $\Delta_{\rm x,cell}$ are the cell gas density, gas vertical velocity and size. We select only the gas that is outflowing: that has a vertical velocity that is positive above the plane of the disc and negative below the plane of the disc. All quantities are very similar in the first 50 Myr, but the outflow rate quickly drops without CR injection or with CR injection but without transport, in the noCR and Advection models. In the Advection model, the lack of CR transport prevents the generation of outflows: the amount of outflowing gas is weaker than in the absence of CR energy, perhaps because the SN bubbles have more difficulty breaking out from the disc that is puffed up by CR pressure. 
The outflow rate in the Streaming simulation is very similar to the Advection case due to the low values of the Alfv\'en velocity compared to the SN shock velocity (see discussion in Section~\ref{section:mag}).
For the Streaming Boost simulation, the outflow rate is similar to the Advection case at early times, but it rises to higher values at later times, although the outflow rate is sill weaker than with the other transport models. 
The mass outflow rate is higher when allowing for CR diffusion: CR energy is allowed to escape high density regions and accelerate more diffuse gas, that is accelerated at higher velocities than dense gas. The isotropic diffusion model and anisotropic diffusion with $\kappa= 3 \times 10^{28} - 1\times 10^{29}\, \rm{cm}^{2} \rm{s}^{-1}$ are the most efficient at driving winds: more that 10 times more gas is driven out of the 2 kpc and 10 kpc planes compared to the noCR case. The mass-averaged outflow velocity is less affected by the injection of CRs: it is increased at most by a factor of 2 for the highest diffusion coefficient, and for the lowest diffusion coefficient, it is even lower than in the noCR model. In the Isodiff run, the amount of CR diffusion is expected to be equivalent to the anisotropic diffusion run with $\kappa = 10^{28}\,\rm{cm}^{2}\,\rm{s}^{-1}$ for a randomly oriented magnetic field. However, the wind mass loading is stronger in the Isodiff run, indicating that field lines in the disc are, indeed, preferentially within the plane of the galaxy suppressing the effective vertical diffusion of CRs (see Section~\ref{section:bfieldevol}). The wind after 250 Myr is $\sim 5$ times denser with CR injection and diffusion with high diffusion coefficients than without CR injection.

\begin{figure*}
  \centering
  \subfloat[Thermal pressure gradient seen edge-on after 250 Myr]{\includegraphics[width=\textwidth]{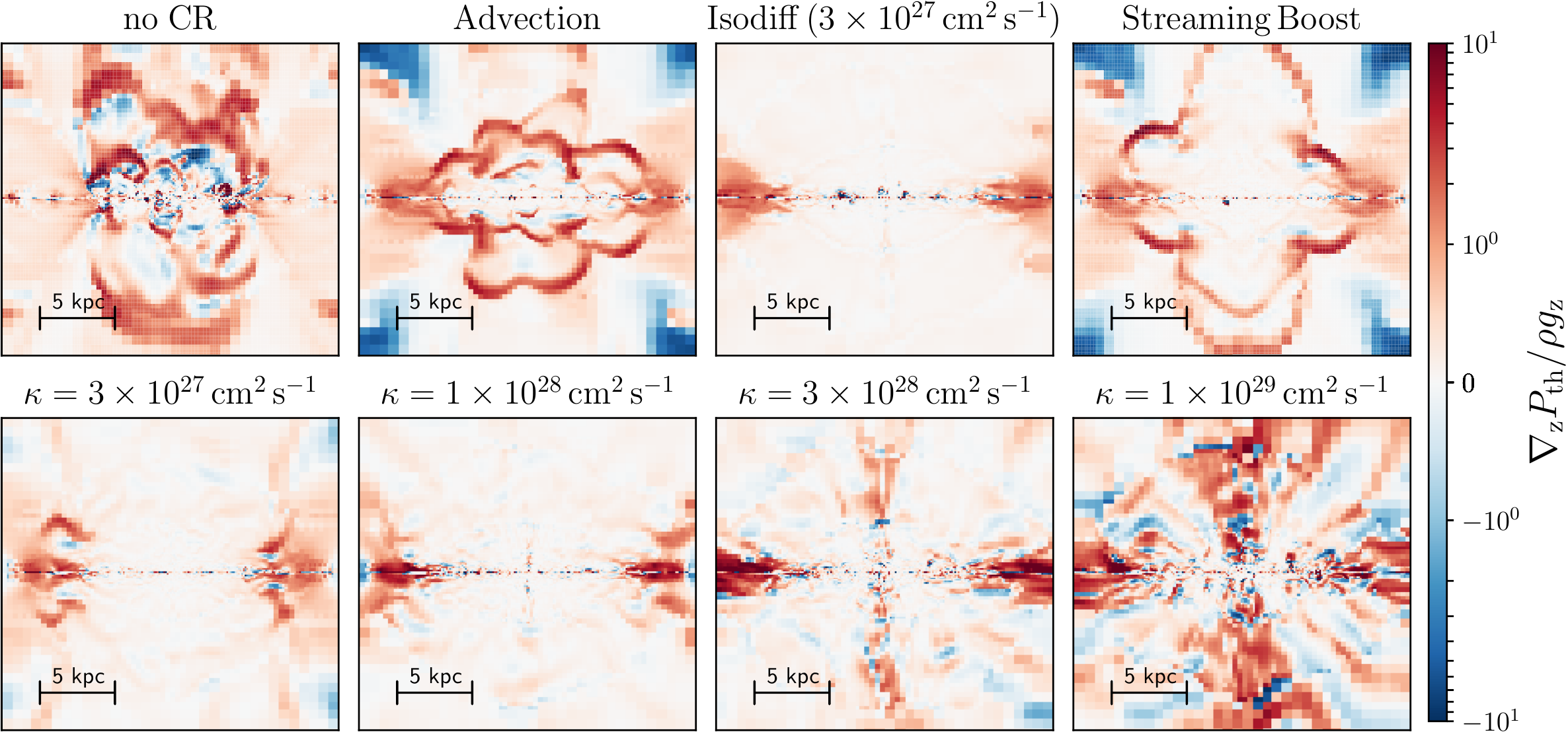}}\\
  \subfloat[CR pressure gradient seen edge-on after 250 Myr ]{\includegraphics[width=\textwidth]{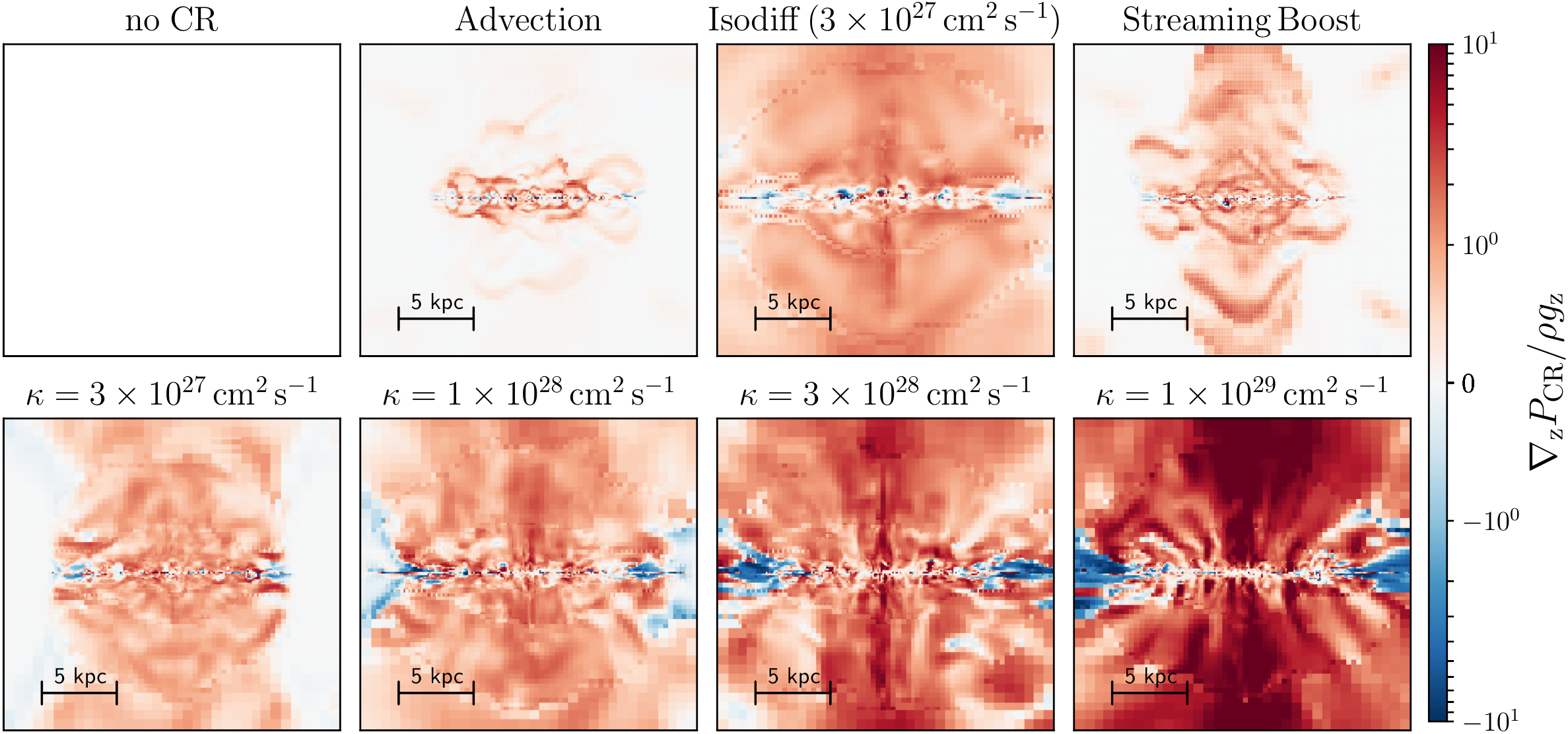}}
  \caption{Thermal pressure gradient (a, two upper rows), and CR pressure gradients (b, two bottom rows) relative to the gravitational vertical pull, in 20 kpc slices of the G9 galaxy viewed edge-on, for the different runs, as indicated on the panels. A positive (red) ratio corresponds to a pressure gradient that pushes outwards. A linear threshold of 1 is used for the symmetric logarithmic colourbar. In the absence of CR injection or transport (noCR and Advection), the pressure above and below the mid-plane is dominated by thermal pressure gradient that is maximal in shocked regions.  When including CR transport, the gradient of the CR pressure is much higher than that of the thermal pressure, and the gradient of the thermal pressure is lower and less shocks form. The ratio of the thermal pressure gradient to gravity exceeds unity only within shocks whereas the CR pressure gradient in the wind becomes uniformly greater (2 to 10 times) than the gravitational pull for $\kappa \geq 3 \times 10^{28}\,\rm{cm}^{2}\rm{s}^{-1}$.}
  \label{Pressure_1kpc}
\end{figure*}

Figure~\ref{Pressure_1kpc} shows edge-on slices of the ratio of the gradient of the thermal (panels a) and CR (panels b) pressures over the vertical gravitational force in the different runs. A positive (red) ratio corresponds to a pressure gradient that pushes outwards. In the absence of CR injection or transport (noCR and Advection), the pressure above and below the mid-plane is dominated by the thermal pressure gradient, which is maximal in shocked regions. When including CR transport, the gradient of the CR pressure is much higher than that of the thermal pressure, and the gradient of the thermal pressure is lower than without CR transport because less shocks form~(see Appendix~\ref{appendix:shock}): the reason is that above and below the plane, the pressure is dominated by CR pressure (as shown in Fig.~\ref{edgeon1}), of which the distribution is very smooth thanks to the transport of CR energy. One sees that with anisotropic diffusion, unlike with isotropic diffusion, the gradient of the thermal pressure is slightly higher in the wind, which stems from the fact that with anisotropic diffusion, CR transport is suppressed in the directions perpendicular to the magnetic field: this slightly mitigates the smoothness of the total pressure, and therefore shocks are more frequent. The ratio of the thermal pressure gradient to gravity exceeds unity only within shocks whereas the CR pressure gradient in the wind becomes uniformly greater (2 to 10 times) than the gravitational pull for $\kappa \geq 3 \times 10^{28}\,\rm{cm}^{2}\rm{s}^{-1}$, and increases with the increasing value of the diffusion coefficient. Our results agree with that of \cite{Girichidis18}~\citep[see also][]{Simpson16}, where it is found that, when including CR diffusion, above $\sim 1$ kpc away from the mid-plane, CRs provide the dominant gas acceleration mechanism. 
Gradients of CR pressure in the wind are higher for larger diffusion coefficient as more CR energy is allowed to escape the disc without hadronic and Coulomb losses as will be discussed in Section~\ref{sub:CRenergy}.
The Streaming boost case shows a moderate amount of CR gradient of pressure in the wind, lower than that of the lowest anisotropic run studied, $\kappa=3\times 10^{27}\,\rm cm^2\, s^{-1}$, though, even in that case, the CR gradient of pressure overwhelms that of thermal pressure.

\begin{figure}
  \centering
 \subfloat{\includegraphics[width=0.47\columnwidth]{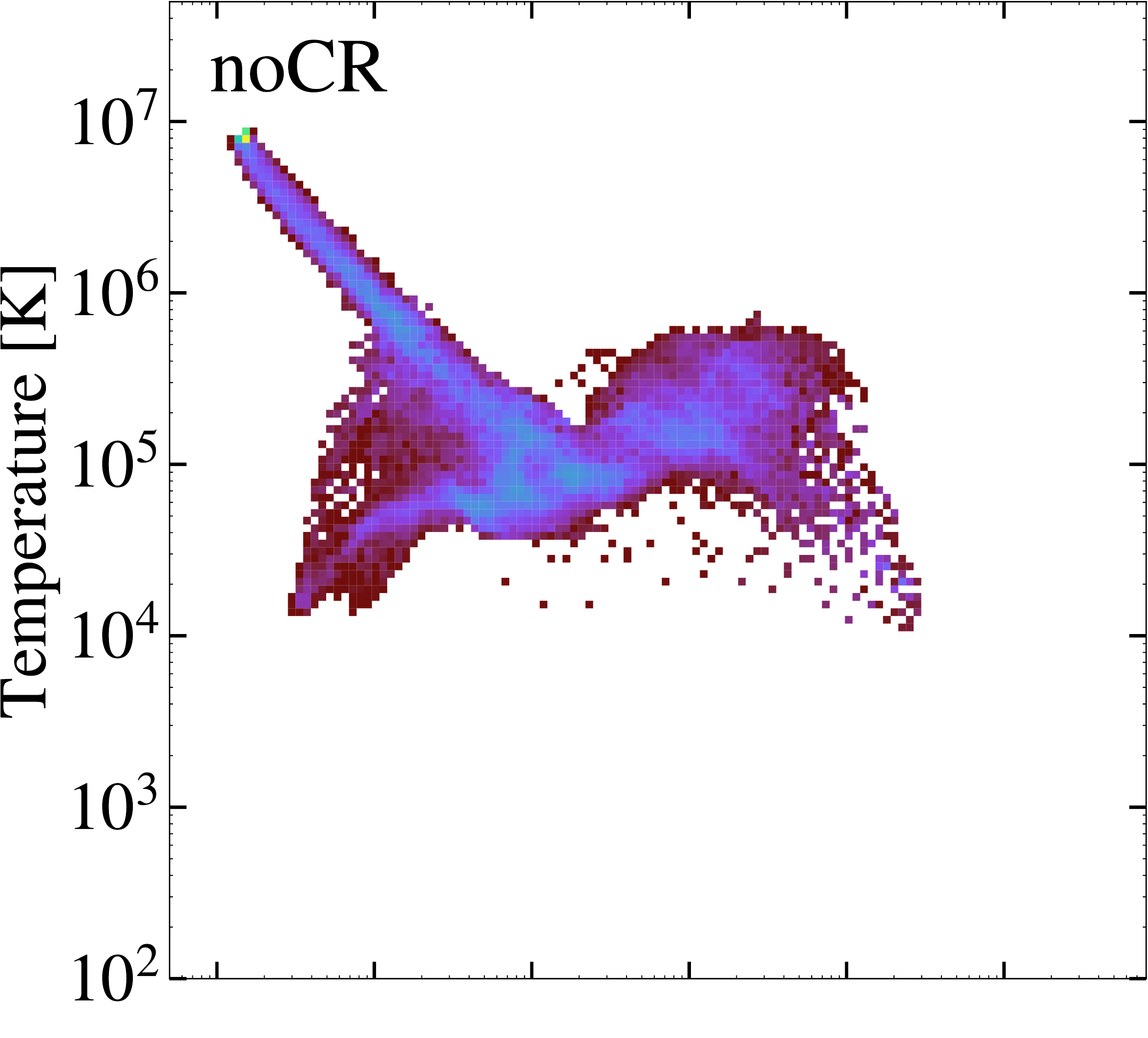}}
\subfloat{\includegraphics[width=0.505\columnwidth]{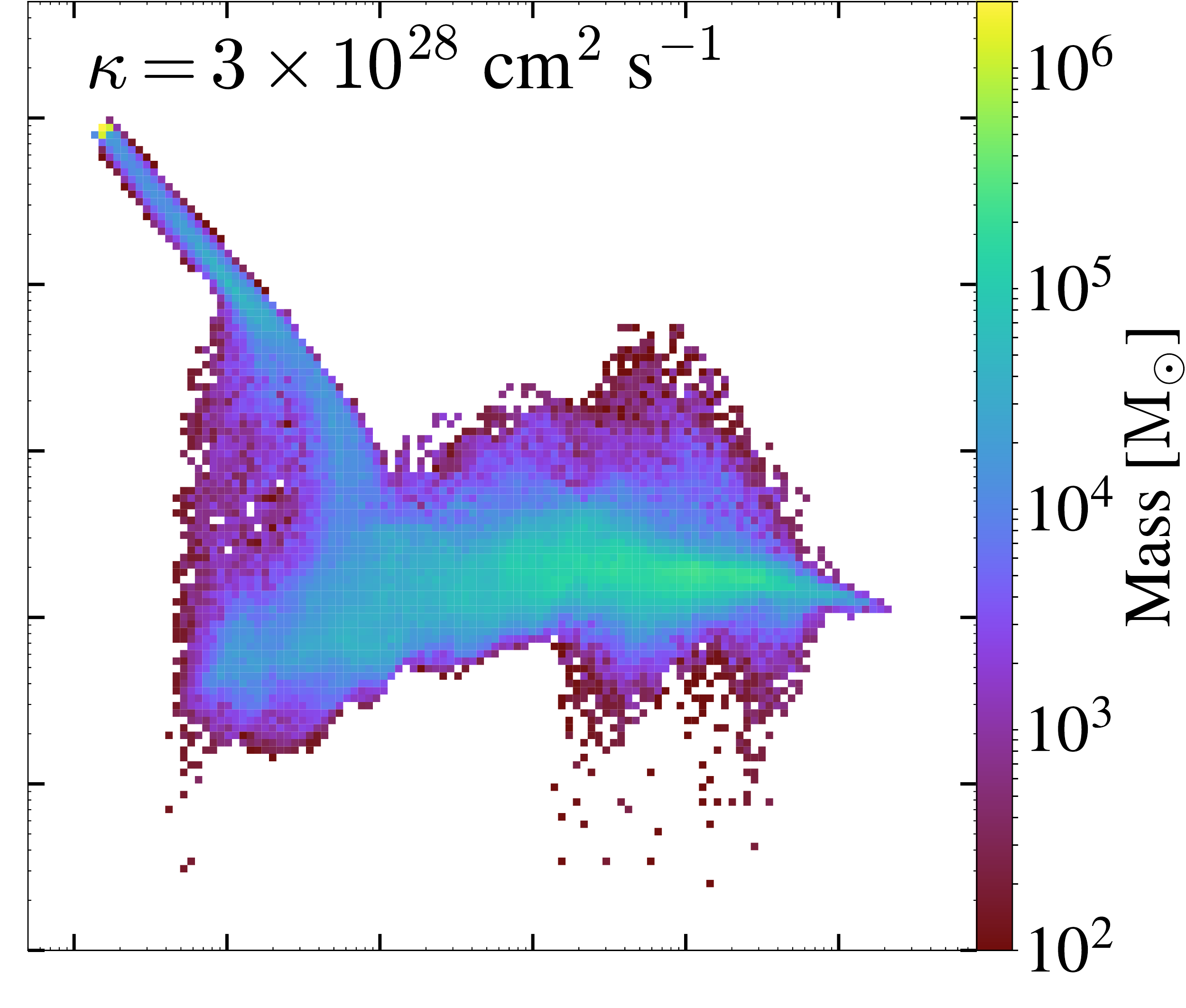}}
\vspace{-14pt}
\subfloat{\includegraphics[width=0.47\columnwidth]{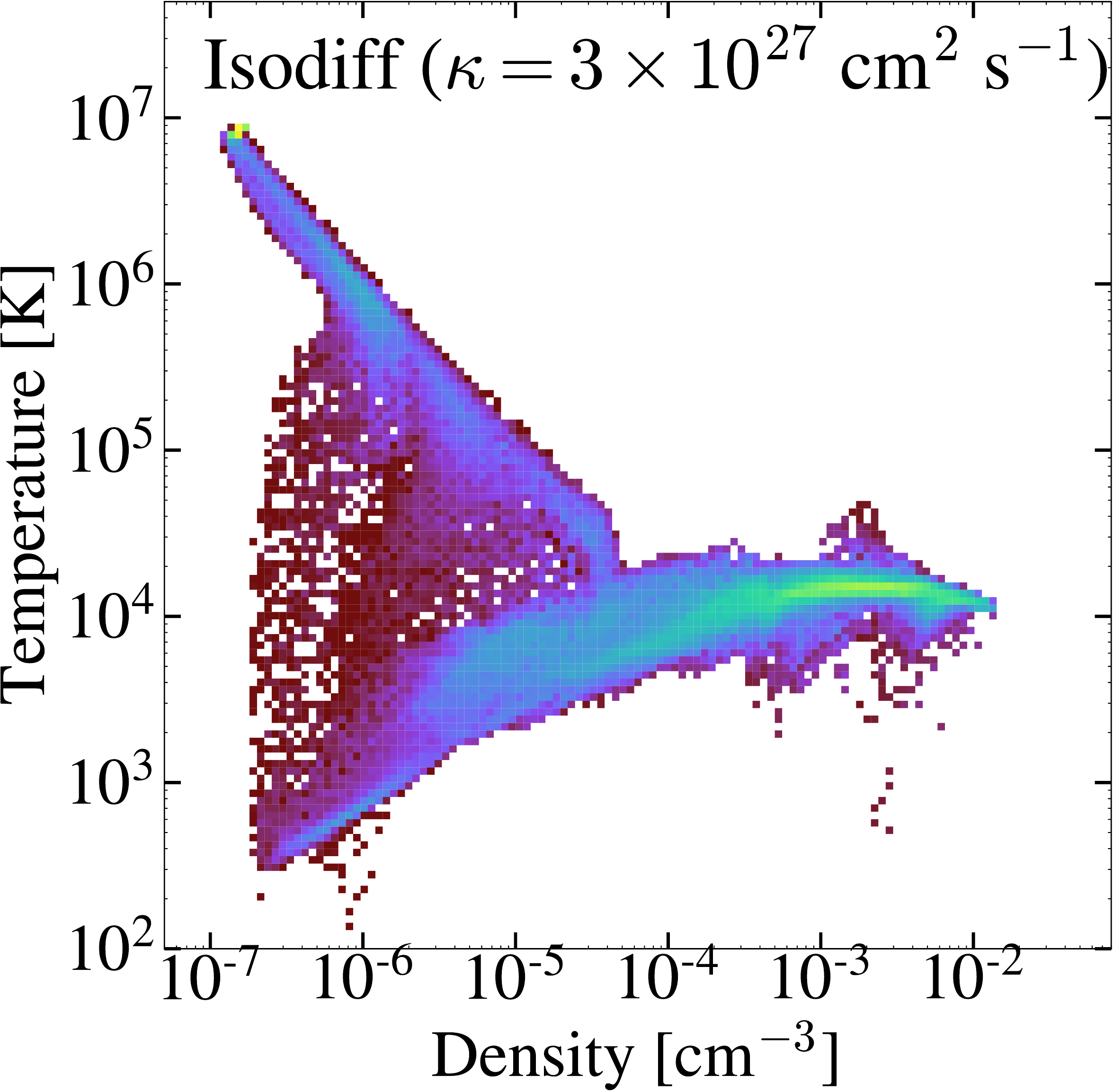}}
 \subfloat{\includegraphics[width=0.505\columnwidth]{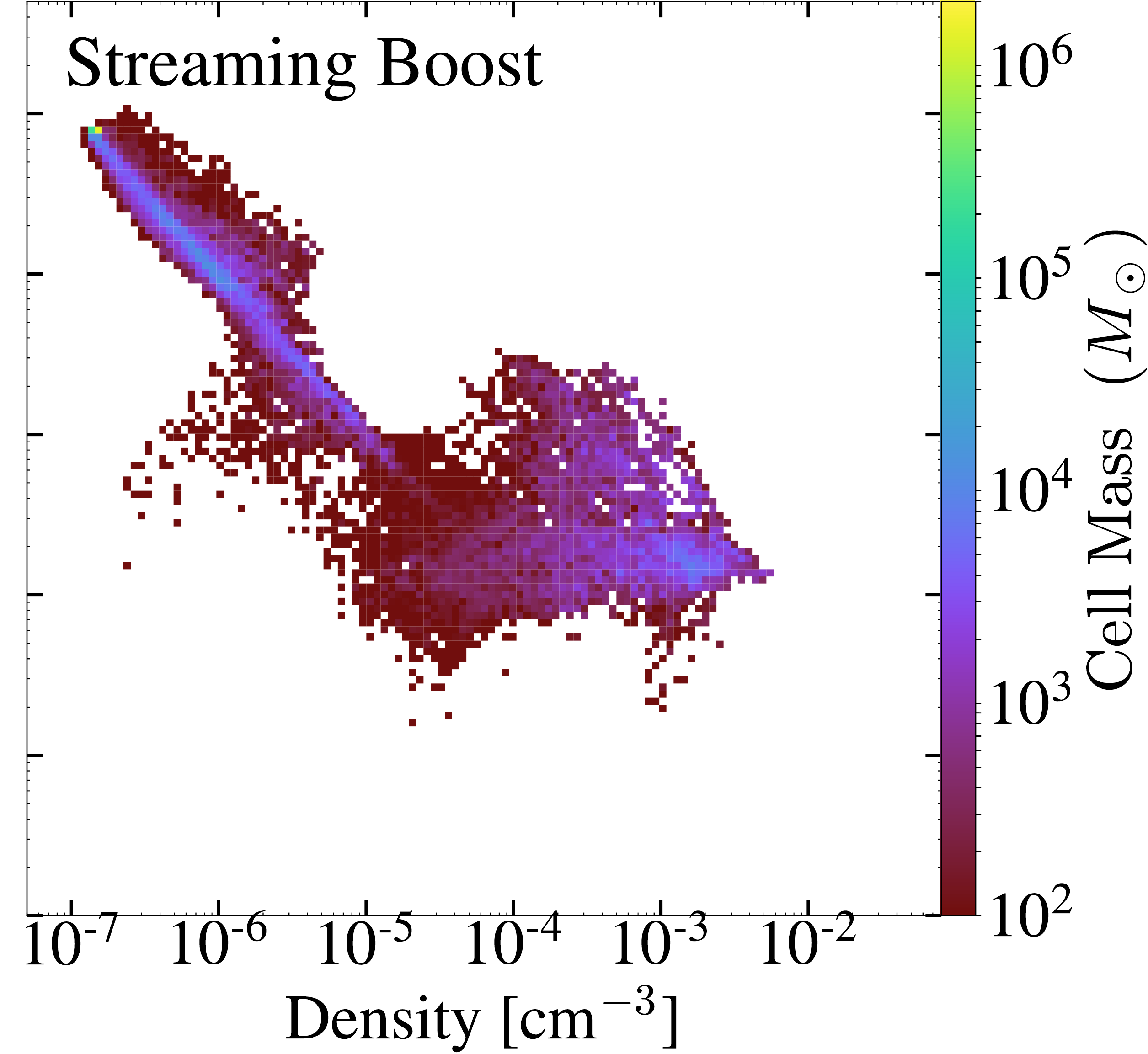}}
  \caption{Total mass of the outflowing gas showed in bins of temperature-density, after 250 Myr for  the noCR, Isodiff, $\kappa=3\times 10^{28}\,\rm cm^2\, s^{-1}$ and Streaming boost runs as indicated in the corresponding panels. We use a lower threshold of $10\, \rm km\, s^{-1}$ to detect the outflowing gas, selected only above 2 kpc from the disc plane. Without CR injection, the densest ($10^{-4}-10^{-3}$ cm$^{-3}$) component of the outflowing gas has temperatures of $10^{5} - 10^{5.5}$ K. When including CR injection with transport, that component is colder ($10^{4}$ K) and denser ($10^{-4}-10^{-2}$ cm$^{-3}$). In the Advection case, the wind component is almost nonexistent. The wind component is very weak with CR streaming but with similar temperatures to simulations with CR diffusion. The upper left part of the phase diagram corresponds to gas of the intergalactic medium at the outer edges of the disc.}
  \label{Phase_diagram}
\end{figure}

Figure~\ref{Phase_diagram} shows the temperature-density phase diagrams in the outflowing gas after 250 Myr for a representative subset of simulations (noCR, Isodiff, $\kappa=3\times 10^{28}\,\rm cm^2\, s^{-1}$ and Streaming boost).  We use a lower threshold of 10 $\rm km\,s^{-1}$ to detect the outflowing gas, selected only above 2 kpc from the disc plane. Without CR injection, the densest ($10^{-4}-10^{-3}$ cm$^{-3}$) component of the outflowing gas has temperatures of $10^{5} - 10^{5.5}$ K. When including CR injection with transport, that component is colder ($10^{4}$ K) and denser ($10^{-4}-10^{-2}$ cm$^{-3}$), in agreement with recent work from \cite{Girichidis18} who simulated detailed slabs of the ISM. Interestingly, the temperature of the wind in the isotropic case is much more uniform than when modelling anisotropic diffusion: again, this is due to the fact that diffusion is suppressed in the directions perpendicular to the magnetic field, which reduces the smoothness of the total pressure, and shocks are more frequent, which complicates the temperature distribution of the outflow compared to the isotropic case. With CR injection but no CR transport (Advection), the wind component is almost non-existent (not shown here), as previously described, and the wind component is very weak for CR streaming with boost but has the same temperatures as with CR diffusion.

\subsection{Fate of the cosmic ray energy}
\label{sub:CRenergy}
\begin{figure}
    \centering
    \includegraphics[width=\columnwidth]{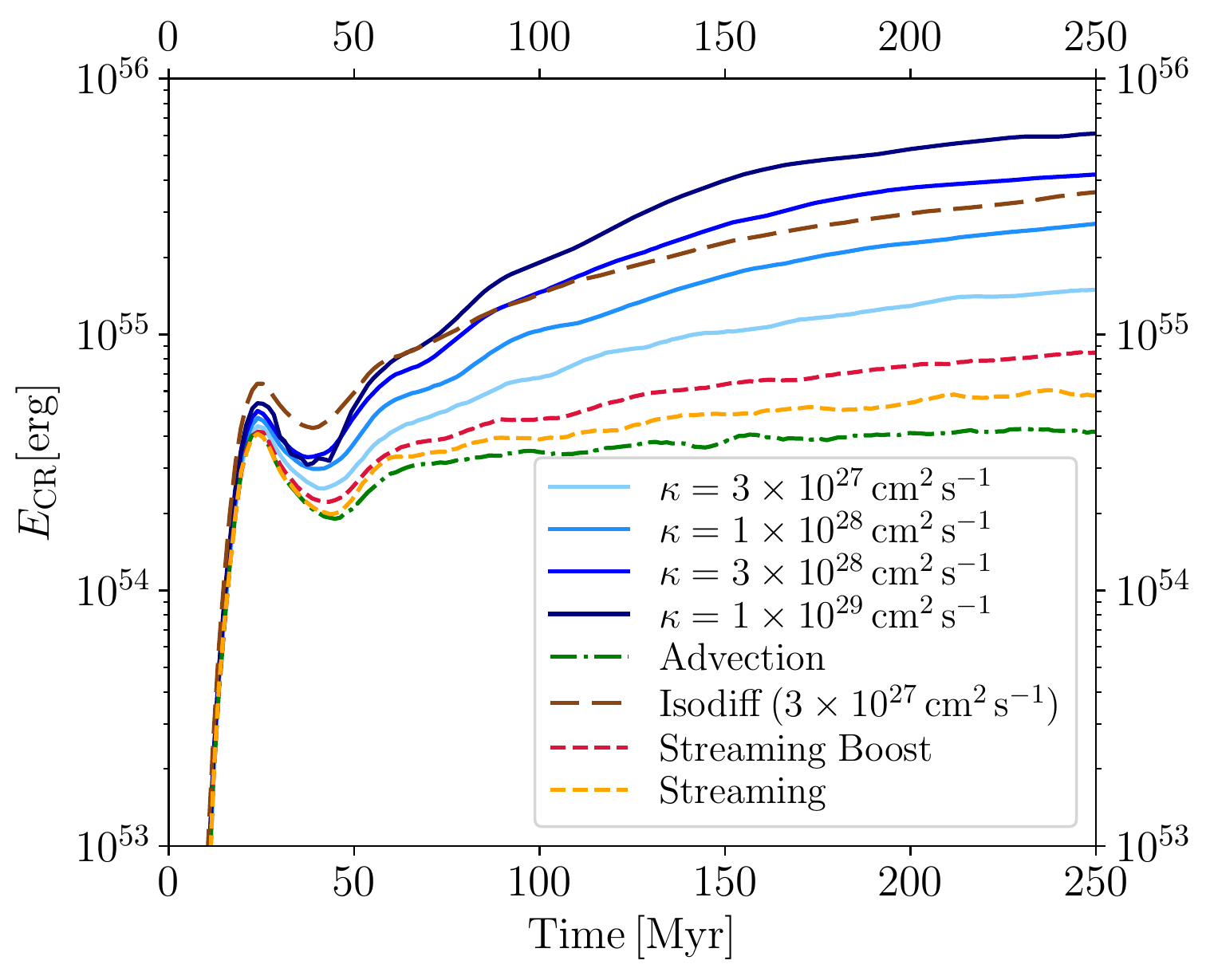}
    \caption{Total CR energy in the box the different runs of the G9 galaxy. The higher the diffusion coefficient, the higher the CR energy: the reason is that the more CRs can diffuse, the more they can escape high gas density and CR-energy-density regions where the CR cooling rate is higher. At the beginning of the simulation, with isotropic diffusion, CR energy quickly escapes in directions perpendicular to the disc, whereas it remains trapped in the disc for anisotropic diffusion.}
    \label{Ecrtot}
\end{figure}

\begin{figure}
    \centering
    \includegraphics[width=\columnwidth]{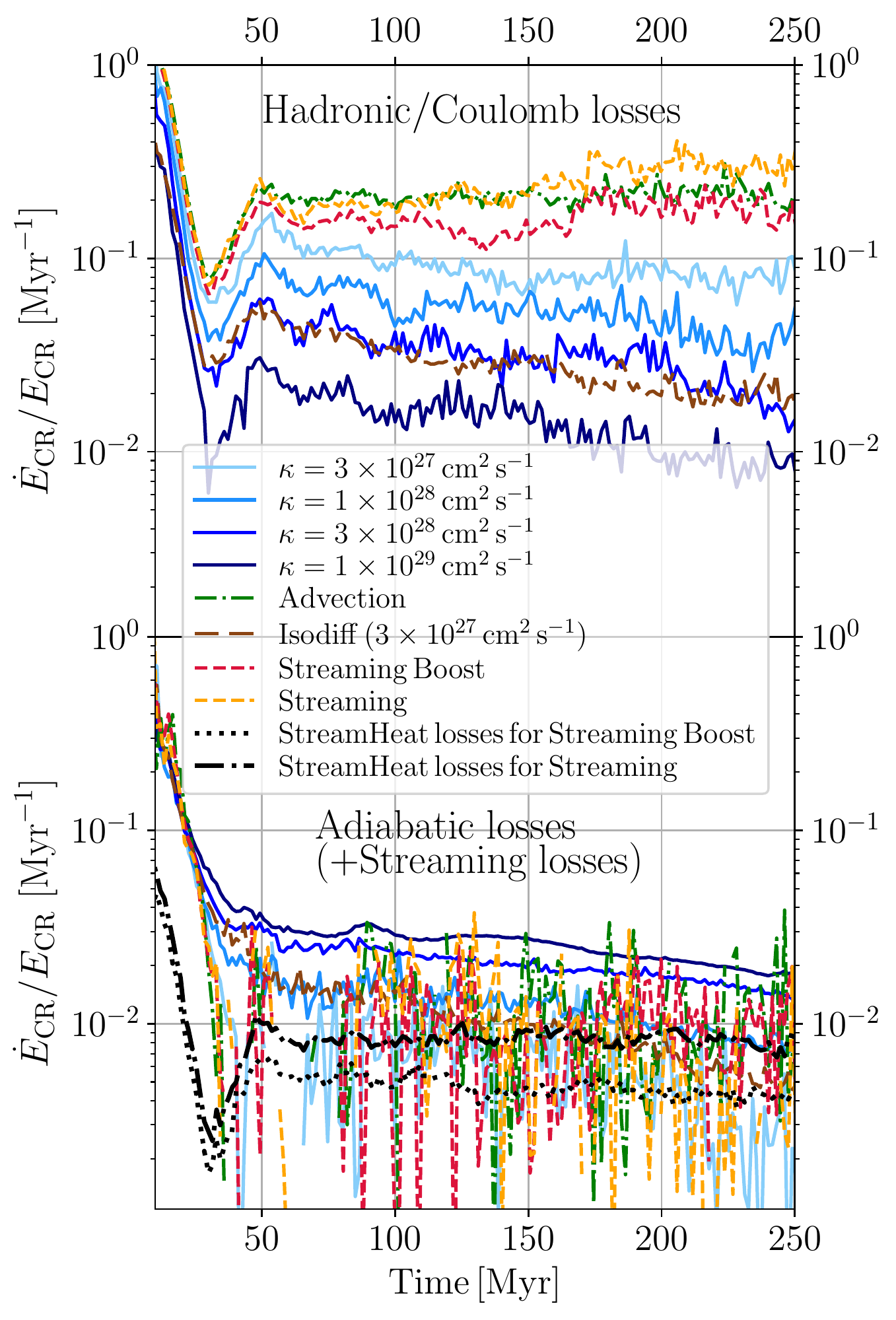}
    \caption{\emph{Top panel:} Hadronic and Coulomb cooling rate of CR energy inside the box for the different runs. \emph{Bottom panel:} CR adiabatic losses ($-P_{\rm CR}\nabla\cdot \mathbf{u}<0$). We find that the cooling rate is higher for lower diffusion. We find that adiabatic losses are subdominant except in the isotropic and anisotropic diffusion simulations with the highest diffusion coefficients. The black dotted line corresponds to the streaming losses ($-\vec{u}_{\rm{st}} \cdot \nabla P_{\rm CR}$) in the `Streaming Boost' simulation.}
    \label{Ecrloss}
\end{figure}

Figure~\ref{Ecrtot} shows the total CR energy in the simulated box for the different runs. The amount of CR energy in the box at a given time increases with the diffusion coefficient, the reason being that it escapes more efficiently the high density star-forming regions where it is injected, and where the cooling rate is higher. The total energy in the Advection run is lower for two reasons: first because the SFR is lower and therefore the total CR injection is weaker, second because the injected CR energy stays confined in the higher density regions from where it can escape only by advection with the gas, and therefore it looses energy to CR cooling (cf. Fig.~\ref{Ecrloss}). In the Isodiff run, CR energy rises faster because it does not have to follow the magnetic field lines, thereby avoiding more rapid cooling in the higher density  disc. Conversely, in the anisotropic diffusion, the energy injected in the disc stays trapped in it for longer times. At later times, however, the amount of CR energy in the anisotropic diffusion simulation with $\kappa=3 \times 10^{28} - 1 \times 10^{29}\rm{cm}^{2} \,\rm{s}^{-1} $ is higher than in the Isodiff run because the SFR in that simulation being higher, CR energy injection is higher too, which compensates for the higher cooling rates.

\begin{figure}
    \centering
    \includegraphics[width=\columnwidth]{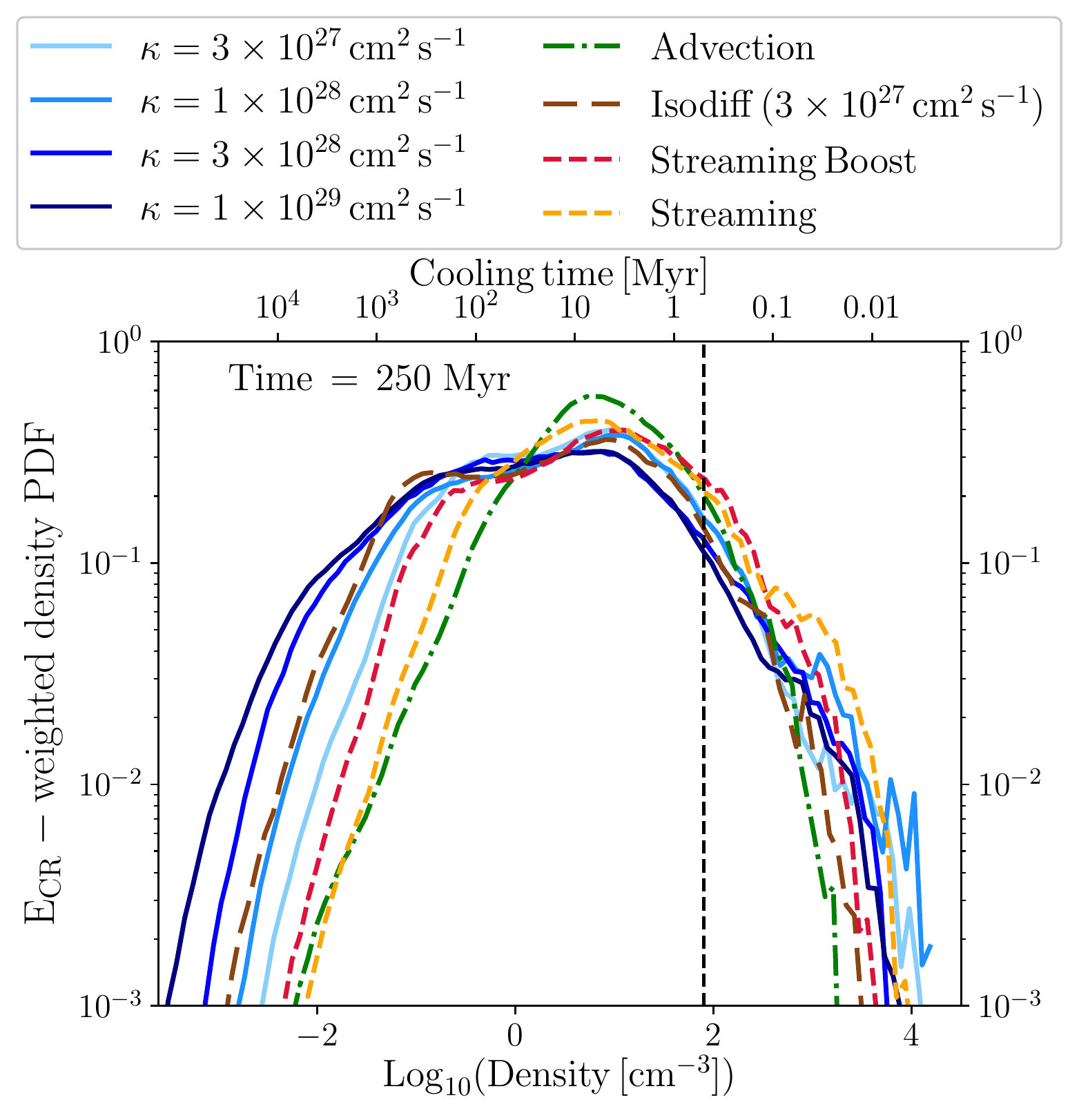}
    \caption{CR energy-weighted density PDF a disc centred on the G9 galaxy (4 kpc height and 10 kpc radius), at 250 Myr. The upper axis shows the corresponding hadronic and coulomb cooling time. The vertical black dashed line is the density threshold for star formation. }
    \label{Ecrpdf}
\end{figure}

Figure~\ref{Ecrloss} displays the sum of hadronic and Coulomb CR energy losses relative to the total energy in the box (upper panel), as well as adiabatic losses\footnote{We found that the adiabatic gain was nonexistent or negligible depending on the run so that we decided not to show it.} (bottom panel), that is CR energy exchanges with the gas via compression and expansion (the volumetric rate of CR adiabatic losses expresses as $-P_{\rm CR} \nabla \cdot \mathbf{u}$). In the bottom panel, we also show the losses due to Alf\'en wave heating in the Streaming boost simulations $-\vec{u}_{\rm{st}} \cdot \nabla P_{\rm CR}$ (recall that the loss term is not boosted). We find that Alfv\'en wave heating is subdominant and that adiabatic losses are mainly subdominant except in the Isodiff and anisotropic diffusion runs with high diffusion coefficient simulations. 
In accordance with previous work \citep{Pfrommer17gray,Girichidis18,Chan2018,Hopkins2019}, we find that the hadronic and Coulomb cooling rate is higher for lower diffusion and the two streaming solutions. 
The reason is that CR energy spends more time in dense regions where the hadronic and Coulomb CR cooling rate is higher (it scales linearly with gas density), as shown in Fig.~\ref{Ecrpdf}, which displays, at different times, the instantaneous PDF of the density in the disc, weighted by CR energy: it shows the densities at which bulk of CR energy lies. One sees that in the Advection model, CR energy is concentrated in regions with densities that are higher than when including diffusion, and CR diffusion allows CR energy to populate lower density regions, where the hadronic and Coulomb cooling time is longer.

\subsection{Magnetic field evolution}
\label{section:bfieldevol}

\begin{figure}
    \includegraphics[width=\columnwidth]{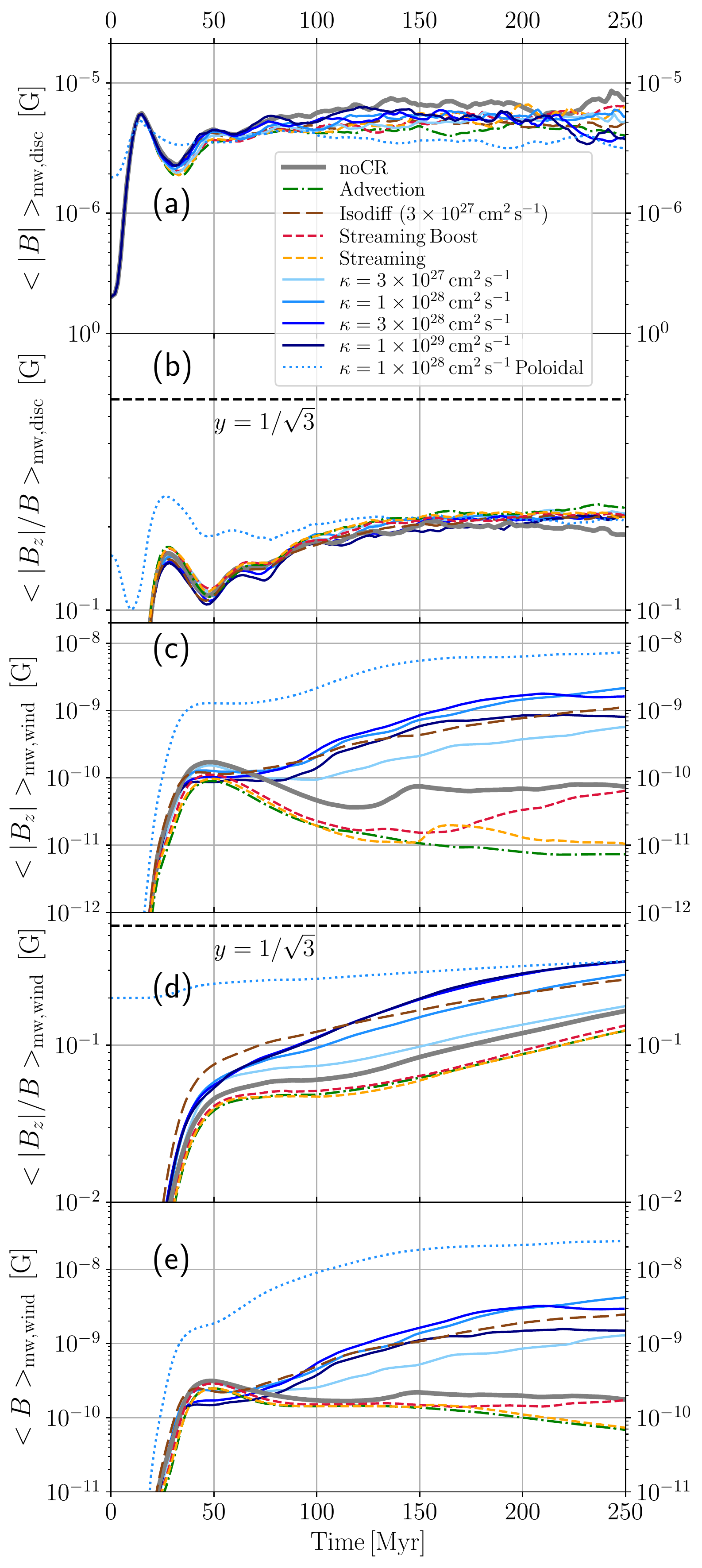}
    \caption{Evolution of the (a) mass  averaged magnetic field  $\vec{B}$ in the disc (of 4 kpc height and 10 kpc radius); (b) mass averaged ratio of $\vec{B}_{\rm z}$ over $\vec{B}$ in the disc (c) mass averaged $\vec{B}_{\rm z}$ in the wind (more than 2 kpc away from the disc) (d) mass averaged ratio of $\vec{B}_{\rm z}$ over $\vec{B}$ in the disc wind; mass average norm of $\vec{B}$ in the wind. The mean magnetic field strength in the disc saturates at $\langle B \rangle\simeq 5\,\mu\rm  G$ with a similar vertical structure in all of the simulated cases. In the wind, a clear difference appears when adding diffusion (isotropic or anisotropic), and the vertical component is more than ten times greater with CR injection and diffusion than without CR injection. }
    \label{Bfield}
\end{figure}

We show in Fig.~\ref{Bfield} the evolution of the vertical component -- perpendicular to the disc -- of the magnetic field for the different runs, in the disc (panels a,b) and in the wind (panels c,d), and the norm of the magnetic field vector in the wind (panel e). This is particularly relevant to CR transport since CR streaming and anisotropic diffusion are along magnetic field lines, and additionally the efficiency of streaming scales with the strength of the magnetic field.
If the vertical component is strong, CR energy can escape from the disc to more diffuse and higher altitude regions. Conversely, if the magnetic field lines are parallel to the disc, CR energy is confined at lower altitudes close to the disc. In the disc (panel a), the amplification of the vertical component of the magnetic field is very similar in the different runs, although the noCR runs stands out with a slightly higher amplification. As shown in panel b, the vertical component is on average 2 -- 3 times lower than the value for a randomly oriented field -- where one would have $1/\sqrt{3} \times |\vec{B}|$ --, which shows that in the disc, the magnetic field lines are dominated by their azimuthal component. Therefore, when the transport is anisotropic, CRs are more confined in the disc because they follow the magnetic field lines, as opposed to isotropic diffusion. This explains the fact that isotropic diffusion with $\kappa=3\times 10^{27} \,\rm{cm}^2\,s^{-1}$ is more efficient at generating winds than the anisotropic run with $\kappa=1\times 10^{28} \,\rm{cm}^2\,s^{-1}$ since field lines are more likely to be within the plane of the disc than along the direction of propagation of the galactic wind.

In the wind (panel c), a clear difference appears when adding diffusion (isotropic or anisotropic): the vertical component is more than ten times greater with CR injection and diffusion than without CR injection. As shown in panel d, this difference is partly due to the fact that the vertical component is greater (relative to the total amplitude of the magnetic field) in the runs with diffusion, but the main reason is that the total amplitude of the magnetic field is greater in the wind, as shown in panel e. One sees that in the anisotropic diffusion runs, the vertical component is roughly 3 to 4 times smaller than the norm of the magnetic field, which means that the field in the wind is also dominated by its azimuthal component: this also explains the fact that isotropic diffusion is more efficient at generating winds than the anisotropic run with $\kappa=10^{28} \,\rm{cm}^2\,s^{-1}$ because more CR energy is transported vertically.
Nonetheless, the magnetic field lines are more vertical in the wind than they are in the disc indicating that the launching of the wind is able to open up the strongly azimuthal field lines from the disc.
The higher values of the magnetic field in the wind in the runs with diffusion come from the largest densities in the wind, as magnetic field scales with density (not shown here, but the scaling of the magnetic field shows a linear dependency with density with a large scatter) in agreement with~\cite{Pakmor16}.

The magnetic field in the disc quickly increases -- as a result of both differential rotation~\citep{DT10}, turbulent dynamo~\citep{RT16} or in combination with CRs~\citep{Hanasz04} --   in the first $100\,\rm Myr$ until it saturates at a value of $<B>_{\rm mw,disc}$ between 3--6 $\mu G$ depending on the simulation.
Those final values of the saturated magnetic field are in agreement with observations~\citep[see e.g.][]{Beck2000}, and can reach values as high as 10--20$\, \mu \rm G$ in the dense star-forming gas, consistent with the magnetic field observed in cold clouds~\citep{Crutcher2012}.
The saturated values of the magnetic obtained in the simulated disc are broadly consistent with those obtained in~\cite{Pakmor16} or~\cite{Pfrommer17} for a similar halo mass, despite the more resolved ISM structure, due to higher resolution, obtained here.
This sheds light onto the delay for the Streaming run to operate: the characteristic wave speed of the streaming instability is the Alfv\'en velocity (here boosted by a factor of 4), and until the magnetic field reaches rough equipartition with the kinetic and thermal energy in the disc, the Alfv\'en velocity cannot compete with others characteristic velocities of the flow. 
It is possible to estimate crudely the expected effect of streaming in comparison to diffusion.
Considering star-forming clouds of size of $L_{\rm c}\simeq 50\rm pc$, the characteristic diffusion velocity is $u_{\rm D}=\kappa/L_{\rm c}$, hence, for $\kappa=3\times10^{27}\,\rm cm^2\,s^{-1}$, we obtain $u_{\rm D}=195\, \rm km\,s^{-1}$. 
The boosted streaming velocity in comparison is of the order $4\times u_{\rm A}=50\, \rm km \, s^{-1}$ for a magnetic field of $5 \, \rm \mu G$ and a gas density of $10\,\rm cm^{-3}$ (in practice the boosted streaming velocity in the ISM varies from a few to $100\, \rm km \, s^{-1}$), well below the velocity of the smallest diffusion coefficient used here.
Indeed, the streaming model in which the streaming velocity equals the Alfv\'en velocity produces galactic-wide properties more similar to that of the Advection case than they already are in the boosted case, that is it further weakens the mass loading of the galactic wind.

\section{Galaxy/Halo mass dependency}
\label{G8/G9}

We perform simulations of a subset of the models presented in Table \ref{tab2} for the smaller G8 galaxy. We use the same spatial resolution of 9 pc, and the same resolution for the formed stellar particle mass, although the mass of the initial stellar particles (that do not explode as SNe) is ten times smaller than for the simulations of the G9 galaxy. The DM particle mass is also ten times smaller. 

\begin{figure}
    \centering
    \includegraphics[width=0.95\columnwidth]{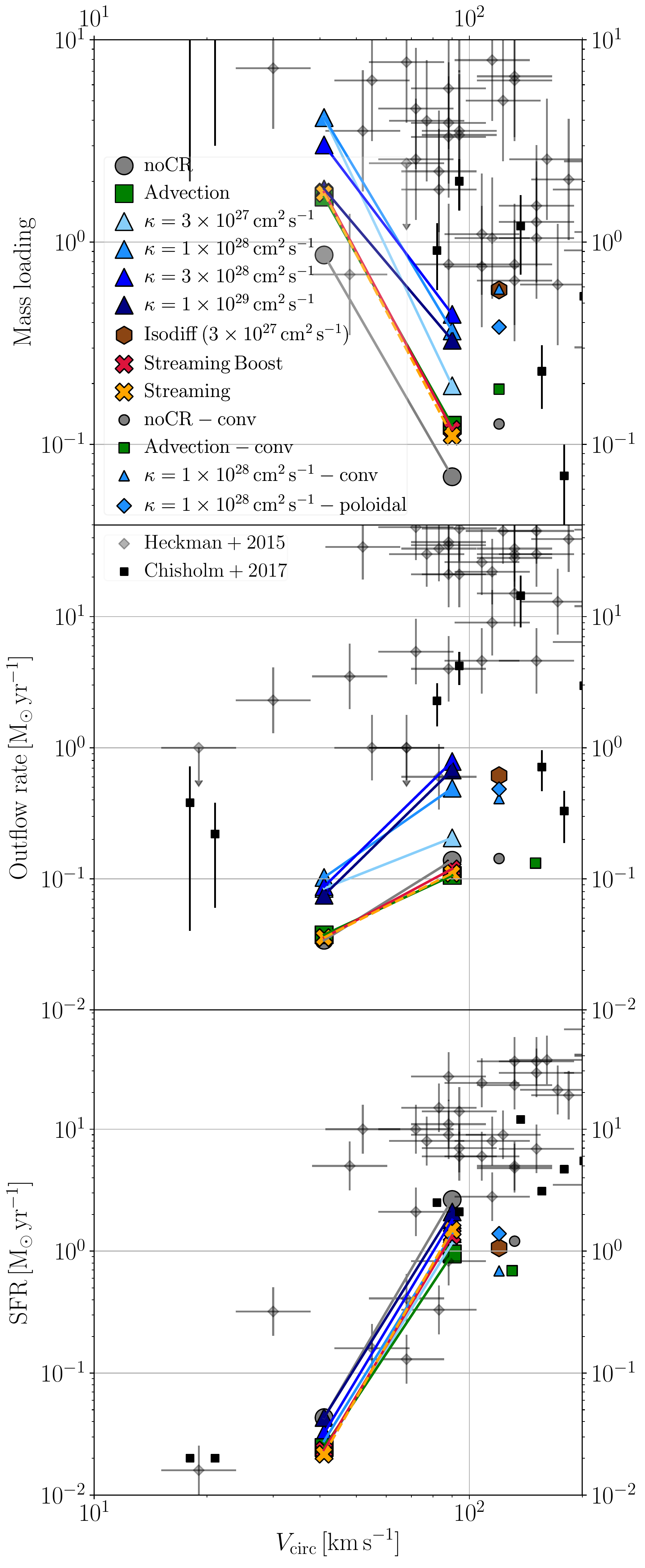}
    \caption{Markers show the time averaged (between 75 -- 250 Myr) outflow rates (top) mass loading factors (middle) and SFR (bottom) as a function of the circular keplerian velocity of G8 (41 $\rm{km}\,{s}^{-1}$) and G9 (90 $\rm{km}\,{s}^{-1}$) at the galactic scale radii. For G9, some of the markers are shifted for clarity. The data points are taken from \cite{Heckman2015} and \cite{Chisholm2017}. The labels with the suffix "-conv" are convergence study runs and are discussed in Section \ref{convergence}. The poloidal (blue diamond) run is discussed in Section \ref{Binit}.}
    \label{scaling}
\end{figure}

In Fig.~\ref{scaling}, we show how the mass-loading (top panel), outflow rate (middle panel), and SFR (bottom panel) scale with the circular velocities of the G8/G9 galaxies and how these scalings compare with recent observations from \cite{Heckman2015} and \cite{Chisholm2017}. 
We compute the circular keplerian velocity at the galactic scale radii of the G8 and G9 galaxies, and we average the outflow rate, mass loading factor and SFR between 75 and 250 Myr.
Overall the SFR of the galaxies for any of the models are very similar and fall within the observed SFR. 
 As a result of CR feedback, the SFR for the low-mass G8 galaxy are reduced by a factor $\sim 1.5-2$ and the impact of CRs on the SFR is reduced with larger diffusion coefficients in a similar fashion to G9.
 Therefore, as in the G9 galaxy, the injection of CRs in the G8 galaxy has a mild impact on star formation.

These observational studies by \cite{Heckman2015} and \cite{Chisholm2017} use UV absorption lines to measure outflow velocities and determine mass loss rates in nearby star-forming galaxies. The inferred mass loading factors are rather uncertain. These quantities can be directly determined in numerical simulations whereas observationally, they rely on challenging measurements and estimates, and caution is therefore needed in the interpretation of this comparison. Namely, the radius of the outflow measurement is rather uncertain. We measure the outflow at 1kpc for the G8 galaxy close to the value of $2 r_{*}$ where $r_{*}$ is the half-light radius of the galaxy, following \cite{Heckman2015}.  
The mass loading factor and outflow rates fall significantly below the observations, especially for the G8 galaxy, despite the significant improvement due to the presence of CR feedback and transport.
Therefore, it suggests that even though the energy injection from the SFR is similar, the outflow generation is still too weak. We note that our results differ from \cite{Jacob18} who find ten times higher mass loading factors for this mass range. This difference might stem from the differences in the initial set-up: \cite{Jacob18} use an initial rotating gaseous halo in a static Hernquist potential, which might produce higher outflow rates -- compared to our simulations where the gas is already settled in a disc -- since the gas initially distributed in a sphere can be swept along with the outflows. However, \cite{Jacob18} also find some disagreement with observations: they find that CR-driven winds drop rapidly with halo mass, much faster than in observations. 
In addition, one should be aware that there are some important uncertainties in the fraction of the CR energy released into SN explosions (or, equivalently, the CR acceleration efficiency at shocks), and, though, our fiducial value of CR energy fraction of 10 per cent is standard, larger values can be considered and would increase the strength of galactic winds~\citep[see][]{Jacob18}.
Another interesting aspect is the different hierarchy of the diffusion coefficient found for G8 and G9. For instance, in the G8 galaxy, the highest diffusion coefficient is the least efficient at launching winds, whereas it is amongst the most efficient ones in G9. Conversely, the lowest diffusion coefficient is the most efficient at driving winds in G8 and the least efficient in G9. This might stem from the fact that if CR energy escapes too quickly from the galaxy, it cannot help generating winds, and at fixed diffusion coefficient, CR energy escapes faster from a smaller size galaxy. 

\section{Convergence study}
\label{convergence}

To study the convergence, we ran a subset of the simulations presented in Table \ref{tab2} of the G9 galaxy with lower spatial resolution. Since the spatial resolution is twice lower, we reduced the density threshold for star formation from 80 cm$^{-1}$ for the higher resolution fiducial runs to 10 cm$^{-1}$ for the lower resolution runs. The masses of the formed stellar particles and the DM particle mass are the same as in our fiducial runs. We show the results in Fig.~\ref{scaling}. The SFRs are slightly higher in the higher resolution runs, and therefore the mass loading factor are slightly higher in the lower resolution run. The values of the outflow rate agree very well for the two resolutions.

\section{Influence of the initial magnetic field topology}
\label{Binit}

We test the influence of the initial magnetic field topology by running one of the models ($\kappa = 1 \times 10^{28} \,\rm{cm^2}\,\rm{s}^{-1}$) with a poloidal topology rather than toroidal. A poloidal $\vec{B}$ field is obtained by setting the value of $\vec{A}$ to 
\begin{align}
    \vec{A}=B_{0}\left(\frac{\rho}{\rho_0}\right)^{2/3} \begin{pmatrix} -y \\ x \\ 0 \end{pmatrix} \,,
\end{align}

\noindent where $x$ and $y$, are the cartesian coordinates, and $\rho$ is the gas density profile as defined in the disc within the vertical and radial cutoffs, $\rho_0$ its normalisation ($\sim 15 \rm{cm}^{-3}$ for both G8 and G9) and $r_{0}$ its scale radius (1.5 kpc for G9 and 0.7 kpc for G8). $B_0$ is set to 1 $\mu$G. We compare the SFR and the outflows in the  $\kappa = 1 \times 10^{28} \,\rm cm^2\,s^{-1}$ simulations run with the two different magnetic field initialisations: toroidal and poloidal. We show the result for the SFR, outflow rate and mass loading in Fig.~\ref{scaling} (blue diamonds). The SFR in the case of poloidal initial topology is similar to the SFR in the case of toroidal initial topology. The outflow rate is slightly higher: the reason is that the vertical component is strong in the initial poloidal topology but nonexistent in the toroidal topology and therefore CR energy escapes more easily in the vertical directions and generates more winds. Fig.~\ref{Bfield} shows the evolution of the vertical component -- perpendicular to the disc -- of the magnetic field for the different runs, in the disc (panels a,b) and in the wind (panels c,d), and the norm of the magnetic field vector in the wind (panel e). In the disc, the vertical component evolution is very similar in all cases. In the wind, the vertical component of the magnetic field is greater with a poloidal initialisation (panel c), because the norm is greater (panel e), but relative to the norm of the magnetic field in the wind, it converges to the same value (panel d).

\section{Summary and discussion}
\label{Conclusion}

We performed high resolution simulations of isolated disc galaxies embedded in halos of $10^{10} \rm{M}_{\odot}$ and $10^{11} \rm{M}_{\odot}$, with and without CR injection, and with different CR transport models. We summarise below our main results:

\begin{itemize}
    \item The amount of star formation is reduced by a factor of $\sim$2, which is due to the suppression of the high density tail of the density PDF by CR pressure support. Overall, the effect of CR injection has a soft effect on star formation.
    \item The injection of CRs significantly increases the wind generation only when diffusive transport is included. The isotropic diffusion model ($\kappa= 3 \times 10^{27}\, \rm{cm}^{2} \rm{s}^{-1}$) and anisotropic diffusion with $\kappa= 3 \times 10^{28} - 1 \times 10^{29}\, \rm{cm}^{2} \rm{s}^{-1}$ are more than 10 times more efficient at driving winds than without CR injection. Outflows with CR diffusion are ten times colder and up to a hundred times denser than without CRs. The gradient of CR pressure provides the dominant gas acceleration mechanism above and below the plane of the disc.
    \item The injection and diffusion of CRs increases the total pressure gradient, which exceeds the gravitational pull by a factor of 2 to 10 in the wind. Conversely, the ratio of the thermal pressure gradient to gravity does not exceed unity in the wind except within shocks.
    \item The energy losses for CRs are higher for lower diffusion. The reason is that CR energy spends more time in dense star-forming regions where the CR cooling rate is higher. Adiabatic losses and gain are mostly subdominant, and are higher for higher diffusion coefficients.
    \item The streaming of CRs only plays a minor role: the star formation, outflow rates and morphology are very similar to the case without transport at all. This suggests that CR streaming is not an efficient mechanism to generate winds. Our results on CR streaming differ from \cite{Ruszkowski17} but agree with \cite{Chan2018} and \cite{Hopkins2019}.
    \item Although the injection and transport of CRs increases the mass outflow rate and brings our simulated mass loading factors closer to observations and in much better agreement than without CRs, the outflow rates in our simulations are still on the weak part of observed outflow rates.
\end{itemize}

Although this idealised set up of initial conditions allows for a good understanding of the physical impact of the modelled physics, such an isolated disc set-up -- meaning that we do not have a dense pre-existing circum-galactic medium nor realistic anisotropic gas accretion -- is a limitation to how well our simulations can capture the rich interactions of galactic winds with their environment. The next step would be to run cosmological zoom-in simulations of CR feedback in dwarf galaxies, that allow to simulate dwarf galaxies with enough resolution to capture the scales relevant to stellar feedback and enough volume to take into account environmental effects and gas inflow. 

There are as additional forms of stellar feedback processes at play in galaxies that we have neglected here, such as the stellar radiation from the UV producing photo-ionised bubbles or from the IR interacting with the dusty gas~\citep{Hopkins14,Rosdahl15}, that need to be combined with the SN release of CRs~\citep[see][]{Chan2018}.
We have also simplified the process of star formation by using a constant star-formation efficiency, while recent models of simulated galaxies now scale the efficiency with turbulent properties of the star-forming gas~\citep{Semenov16,Kimm17,trebitsch18,Lupi18} calibrated on the multi-free fall models of~\cite{Federrath12}~\citep[based on][]{KM05,PN11,HC11}.
Such coupling of varying star-formation efficiency with the release of CRs in SNe remains to be explored.

Another shortcoming of our model is that the adopted CR physics model is simplified since it considers one single CR momentum-energy bin, and, hence, uniform energy losses and a single diffusion coefficient across the entire spectrum of CR momenta. However, in reality the diffusion coefficient and the amount of energy losses through hadronic and Coulomb losses depend on the considered CR energy. Future improvements will include several CR energy bins to account for the CR spectrum, especially since it represents a crucial observable that one can use in order to constrain wind models \citep{Recchia2017}.

Finally, the injection of CRs at shocks, with CR acceleration efficiencies depending on the structure of the background magnetic field~\citep{Caprioli14}, might change how CRs are released into the ISM~\citep{Pais18,Dubois19}, and, therefore, how they propel the large-scale galactic winds. We defer this to future investigations.

\section*{Acknowledgements}
We thank Corentin Cadiou for his tremendous help with YT and other fruitful scientific discussions. We thank Joakim Rosdahl for his help in setting up the initial conditions and valuable advice and comments on the paper. We thank A. Marcowith for enlightening discussions. This work was granted access to the HPC resources under the allocation A0060406955. This work has made use of the Horizon Cluster hosted by the Institut d'Astrophysique de Paris. We thank S. Rouberol for running smoothly this cluster for us.
%%%%%%%%%%%%%%%%%%%%%%%%%%%%%%%%%%%%%%%%%%%%%%%%%%

%%%%%%%%%%%%%%%%%%%% REFERENCES %%%%%%%%%%%%%%%%%%

% The best way to enter references is to use BibTeX:

\bibliographystyle{aa}
\bibliography{bibliography} % if your bibtex file is called example.bib

%%%%%%%%%%%%%%%%% APPENDICES %%%%%%%%%%%%%%%%%%%%%

\appendix

\section{Shocks galactic winds}
\label{appendix:shock}

\begin{figure}
    \centering
    \includegraphics[width=\columnwidth]{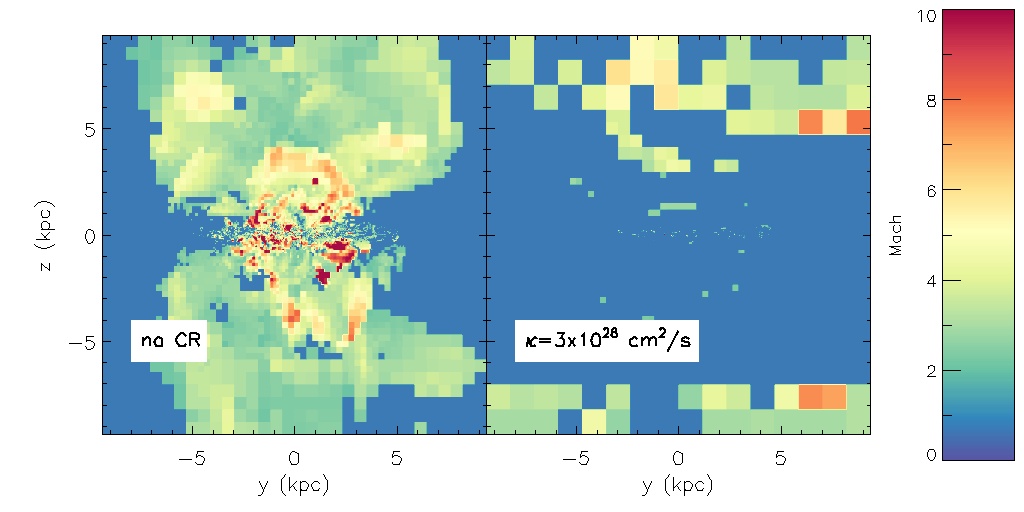}
    \caption{Edge-on view of the maximum value of the Mach number along the line of sight for the simulation without CRs (left panel) and the simulation with CR and anisotropic diffusion with $\kappa=3\times 10^{28}\,\rm cm^2\, s^{-1}$ (right panel) for the G9 simulation at time $t=250\,\rm Myr$.}
    \label{mach}
\end{figure}

We measure the shock Mach number $\mathcal{M}$ in the simulations following the procedure described in~\cite{Dubois19}, where the Mach number is given by the modified Rankine-Hugoniot jump relation for a composite CR-thermal mixture~\citep{Pfrommer17}
\begin{equation}
\mathcal M^2=\frac{1}{\gamma_{\rm e}} \frac{\mathcal{R}_{\rm P}\mathcal{C}}{\mathcal{C}-\left[(\gamma_1+1)+(\gamma_1-1)\mathcal{R}_{\rm P}\right](\gamma_2-1)} \, ,
\end{equation}
where $\mathcal{R}_{\rm P}=P_2/P_1$ is the ratio of the total (thermal plus CR) downstream to upstream pressures (respectively indexed 2 and 1), $\mathcal{C}=[(\gamma_2+1)\mathcal{R}_{\rm P}+(\gamma_2-1)](\gamma_1-1)$, $\gamma_i=P_i/e_i+1$ for $i=\{1,2\}$ and $\gamma_{\rm e}= (\gamma P_{\rm th,2}+ \gamma_{\rm CR} P_{\rm CR,2})/P_2$ for the downstream region. 
Figure~\ref{mach} compares the Mach number for the simulations without CRs and with CR diffusion ($\kappa=3\times 10^{28}\,\rm cm^2\, s^{-1}$). Without CRs, strong shock develop in the immediate circum-galactic medium ($\mathcal{M}>10$) of the galaxy and in the large-scale galactic wind ($\mathcal{M}\simeq4$). With CRs and their diffusive transport, most of those large-scale shocks are erased due to the smooth redistribution of the overall dominating CR pressure.

%%%%%%%%%%%%%%%%%%%%%%%%%%%%%%%%%%%%%%%%%%%%%%%%%%

% Don't change these lines
%\bsp	% typesetting comment
%\label{lastpage}
\end{document}